\documentclass[12pt,preprint]{aastex}

\usepackage{graphics}

\usepackage{amsmath}

\usepackage{natbib}

\newcommand{\be}{\begin{equation}} 
\newcommand{\ee}{\end{equation}}
\newcommand{\bq}{\begin{eqnarray}} 
\newcommand{\eq}{\end{eqnarray}}

\begin{document}

\title{Burning of an hadronic star into a quark or a 
hybrid star}

\author{
Alessandro Drago\altaffilmark{1}, 
Andrea Lavagno\altaffilmark{2} and
Irene Parenti\altaffilmark{1}}

\altaffiltext{1}{Dip. di Fisica, Univ. di Ferrara and INFN
Sez. di Ferrara, I-44100 Ferrara, Italy}
\altaffiltext{2}{Dip. di Fisica, Politecnico di Torino and INFN
Sez. di Torino, I-10129 Torino, Italy}

%\date{\today } 

\begin{abstract} 
We study the hydrodynamical transition from an hadronic star into a
quark or a hybrid star. We discuss the possible mode of burning, using
a fully relativistic formalism and realistic Equations of State in
which hyperons can be present.  We take into account the possibility
that quarks form a diquark condensate.  We also discuss the formation
of a mixed phase of hadrons and quarks, and we indicate which region
of the star can rapidly convert in various possible scenarios.  An
estimate of the final temperature of the system is provided.  We find
that the conversion process always corresponds to a 
deflagration and never to a detonation.  Hydrodynamical instabilities
can develop on the front.  We estimate the increase in the
conversion's velocity due to the formation of wrinkles and we find
that, although the increase is significant, it is not sufficient to
transform the deflagration into a detonation in essentially all
realistic scenarios.  Concerning convection, it does not always
develop. In particular the system does not develop convection if
hyperons are not present in the initial phase and if the newly formed
quark phase is made of ungapped (or weakly gapped) quarks. At the
contrary, the process of conversion from ungapped quark matter to
gapped quarks always allows the formation of a convective layer.
Finally, we discuss possible astrophysical implications of our
results.
\end{abstract}

%\pacs{04.30.Db, 25.75.Nq, 26.60.+c, 97.60.Jd}

\noindent

%\maketitle 


\section{Introduction}

The possible existence of compact stars partially or totally made of
quarks has been proposed many years ago
\citep{witten,Bodmer:1971we,Itoh:1970uw}.  A likely origin of these
compact stellar objects is the conversion of a purely nucleonic (or
hadronic) star into a star containing deconfined quark matter through
a quark deconfinement phase transition.  Several works have discussed
when, during the life of the compact star, the deconfinement
transition should likely take place, either soon after the formation
of the neutron star in the supernova explosion
\citep{Lugones97,Benvenuto99}, or during the cooling of the
protoneutron star \citep{Pons:2001ar}. The formation of quark matter
could also be delayed, if the deconfinement process takes place
through a first order transition so that the purely hadronic star can
spend some time as a metastable object
\citep{Berezhiani:2002ks,Drago:2004vu,Bombaci:2004mt}.

A rather controversial question concerns the duration of the
deconfinement process itself.  In the first paper
discussing this problem \citep{Olinto86} it was assumed that the
conversion proceeds as a slow combustion and it was
found that the velocities involved strongly depend on the
temperature of the star and they can be small if the star is hot.
Soon after the seminal work of Olinto, \citet{Horvath88} studied the
stability of the process and they found that in the presence of
gravity the combustion front becomes instable.  These authors also
stressed that since the appearance of instabilities increases the
velocity of the combustion, it is possible that the slow combustion
becomes a detonation.  On the other hand, no estimate of the velocity
of the conversion front, taking into account the instability, was
presented and the fluidodynamics equations were written in a
non-relativistic frame.  The first relativistic calculation of the
conversion process was done by \citet{Cho93}.  To determine which type
of conversion takes place, either a detonation or a deflagration, they
studied the conservation of the energy-momentum tensor and of the
baryonic flux through the conversion front.  Using very simple
Equations of State (EoSs) for hadronic matter (the Bethe-Johnson and
the Fermi-Dirac EoSs) and the MIT-bag model for quark matter (and
without considering the possibility of a mixed phase) they found that
the conversion is never a detonation and only for very special values
of the parameters it is a slow combustion.  For nearly all the
parameters' values they obtained an unstable front.  In the same
relativistic scheme \citet{Lugones94} studied the case in which the
combustion front is preceded by a precompression wave. In this scheme
they fixed the velocity of combustion and used the conservation law to
determine the final temperature of the combusted phase. Also in this
case (as for \citet{Cho93}) the combusted phase was assumed to be pure
strange matter in $\beta$-equilibrium, described by the MIT-bag model
and it was not considered the possibility of a mixed phase.  In an
other work \citet{Lugones02} found that Rayleigh-Taylor instability can 
significantly increase the velocity of the combustion; moreover, the 
presence of a magnetic field can generate a strong asymmetry in the 
propagation flame, with the maximum velocity in the polar direction.

In our investigation of the conversion process we assume that the
formation of quark matter (QM) takes place inside a relatively cold and
$\beta$-stable compact star.  This scenario is compatible both with
the works in which quark deconfinement takes place in a protoneutron
star immediately after deleptonization \citep{Pons:2001ar}, and also
with the possibility of a delayed formation of QM, as discussed in
\citet{Berezhiani:2002ks,Drago:2004vu,Bombaci:2004mt,vidana:2005}.  We
will study the conversion problem from the fluidodynamical point of
view \citep{landau,Cho93}, so that it will be possible to determine
the type of conversion from nucleonic (or hadronic) matter to QM.
Following \citet{landau} there are four possibilities: detonation,
weak detonation, deflagration (or slow combustion) and strong
deflagration.  The fastest process is detonation, and in that case the
front velocity is suprasonic. It is therefore interesting to explore
the possibility that, at least in the case of detonation, the
conversion process is so rapid that only strong interactions can take
place near the front during the conversion, while weak interactions
will restore $\beta$-equilibrium only after the front. The idea to
distinguish between fast processes (typically mediated by the strong
interaction) and slow processes (due to weak interaction) is common to
other physical situations, as, e.g. the computation of viscosity (see
\citet{Lindblom:2001hd,Drago:2003wg}).  In the literature it has
been shown that detonation is difficult to achieve when realistic EoSs
are used and matter is assumed to reach $\beta$-equilibrium during the
conversion process \citep{Cho93}. Since the EoS of matter is stiffer
if $\beta$-processes are forbidden, then, in principle, it could be
easier to obtain a detonation assuming that matter immediately after
the front is not yet in $\beta$-equilibrium.  In this paper we will
consider both the possibility that slow weak interactions take place
only after the conversion to QM (which is due to rapid and flavor
conserving strong interactions) and also the situation in which
$\beta$-equilibrium is immediately restored.  
It is also important to note that neutrino trapping delays
$\beta$-stability. Although in this paper we will not discuss
neutrino trapping, an estimate of its importance can be obtained
comparing the $\beta$-stable with the not $\beta$-stable scenario.
Concerning temperature,
one also need to check both possibilities, namely that the detonation
front is so rapid that matter immediately after the conversion front
is still cold, and the possibility that strong interactions have
enough time to thermalize the newly formed phase.  The situation is
rather different when analyzing deflagration, which is a subsonic
process. Since for deflagration heat
transmission from the burned to the unburned zone can be crucial, one must 
also consider
burned matter at finite temperature.

The main aim of our work is to study the deconfinement process using
realistic EoSs, in particular we will take into account the
possibility of forming a mixed phase of quarks and hadrons and we will
also discuss the effect of the formation of a diquark condensate.
Another important open question that we will investigate is the
possibility for convection to develop. As we will show, although the
conversion front turns out to be always unstable to gravity-induced 
Rayleigh-Taylor instability, convection can
actually develop only for specific choices of the EoSs, for instance
the presence of hyperons in the hadronic phase or the formation of a
diquark condensate in the quark phase will help the formation of a
convective layer.

The outline of our paper is the following: in Sec.~\ref{fluid-eqs} we
discuss the fluidodynamics of the conversion process; in
Sec.~\ref{eos} we present the EoSs used both for hadronic and for QM
and the structure of the resulting compact star.
In this section we also discuss the thermal formation of mixed-phase; in
Sec.~\ref{temperature} we estimate the temperature reached by the
quark (or mixed) phase after deconfinement; in Sec.~\ref{deflagration}
we show the results of our analysis aiming at classifying the type of
conversion; 
in Sec.~\ref{instability} we discuss hydrodynamical instabilities
and their effect on the conversion velocity;
in Sec.~\ref{convection} we discuss the conditions
allowing convection to take place and, finally, in Sec.~\ref{results}
we summarize our findings and discuss the astrophysical implications
of our work.

\section{Fluidodynamics equations of the conversion process}
\label{fluid-eqs}

The starting point of our analysis is given by the
fluidodynamics equations describing the conservation of the
energy-momentum tensor and of the baryon flux density across the
conversion front. Following \citet{landau}, the continuity 
equations for the energy-momentum tensor,
in the frame in which the front is at rest, read:
\bq
(e_h+p_h)v_h \gamma_h^2 &=& (e_q+p_q)v_q \gamma_q^2\label{ene}\, ,\\
(e_h+p_h)v_h^2 \gamma_h^2+p_h&=&(e_q+p_q)v_q^2 \gamma_q^2+p_q\, ,\label{mom}
\eq
while the continuity of the baryon flux density reads:
\be
\rho_B^h v_h\gamma_h=\rho_B^q v_q\gamma_q\, .\label{baryon}
\ee
In the previous equations $e$ is the energy density, $p$ is the
pressure, $\rho_B$ is the baryon density (all in the rest frame);
$v$ is the velocity of a given phase (in the front 
frame)\footnote{Here and in the 
following we assume the velocity of light $c=1$.}
and finally $\gamma$ is the Lorentz factor.
We use the labels $h$ and $q$ to indicate the hadron and the
quark phase, respectively.
From Eqs.~(\ref{ene},\ref{mom}) the velocities of the two phases can
be obtained and they read:
\bq
v_h^2&=&\frac{(p_h-p_q)(e_q+p_h)}{(e_h-e_q)(e_h+p_q)}\label{vh}\, ,\\
v_q^2&=&\frac{(p_q-p_h)(e_h+p_q)}{(e_q-e_h)(e_q+p_h)}\label{vq}\, .
\eq\\
Obviously, the velocity of the hadronic
phase in the front frame $v_h$ equals, but for the sign, 
the velocity of the front in the frame in which the
hadronic phase is at rest.
It is possible to classify the various conversion mechanisms
by comparing the velocities of the phases (in the front frame)
with the corresponding sound velocities $v_s$. 
The conditions are:
\bq
v_h>v_{sh}&v_q<v_{sq}&\text{strong detonation}    \label{strongdet}  \\
v_h>v_{sh}&v_q>v_{sq}&\text{weak detonation}       \label{weakdet}      \\
v_h<v_{sh}&v_q<v_{sq}&\text{weak deflagration}     \label{weakdef}        \\
v_h<v_{sh}&v_q>v_{sq}&\text{strong deflagration}.
\label{strongdef}
\eq 
It is important to remark that the first condition, namely
the relation between the velocity of the hadron phase and 
the sound velocity of the same phase, can be also
replaced by one of the following conditions:
\be 
v_h>v_{sh} \,\, \, \equiv \, \, \, p_h<p_q \, \, \, \equiv \, \, \, v_h>v_q\, ,
\label{pressure-condition}
\ee
and similarly for the reversed condition.
Let us recall that: 
\begin{itemize}
\item strong detonation is the process in which
a shock wave is responsible for the activation of the burning
(this corresponds to what is generally called all simply detonation);
\item weak detonation is generally considered impossible to
realize in a physical system, because it would imply a heat
transmission faster than the velocity of sound \citep{landau};
\item weak deflagration is a process in which heat transport is 
responsible for activating the combustion (this is what is 
generally called combustion);
\item strong deflagration
is an unstable process in which the conversion front cannot be
described as a simple surface, because instabilities
easily take place.
\end{itemize}
In Sec.~\ref{deflagration} we will classify the deconfinement 
process following the above outlined scheme. 

It is important to note that in Eqs.~(\ref{ene},\ref{mom},\ref{baryon})
EoSs at finite temperature have to be used. While in these
equations the temperature appear as a parameter, it is 
possible to relate its value to the heat produced by the
deconfinement transition, as discussed in Sec.~\ref{temperature}.

A remark is in order.
Concerning the {\it actual} velocity of the conversion front,
it can be determined by solving the fluidodynamics equations 
only if the conversion is a detonation. If it is a deflagration,
other physical scales determine 
the velocity (significantly reducing its value), as e.g. heat diffusion
or the production and diffusion of strangeness.
Moreover, if the front is unstable 
the formation of wrinkles on the front surface or the existence of a 
convective layer can significantly increase the velocity. In the 
following we will call deflagrative velocity $v_{df}$ the velocity 
obtained solving Eqs.~(\ref{ene},\ref{mom},\ref{baryon}).

\section{Equations of state of hadrons and quarks \label{eos}}

In this Section we introduce the EoSs that we are going to use in our
calculations, both for the hadronic and the quark phase.  We discuss
the structure of the mixed phase, also taking into account the
possibility that $\beta$-equilibrium is not immediately reached after
deconfinement.  Finally, we present the structure of compact stars
obtained using the previous EoSs.

In this Section for simplicity we will discuss EoSs at $T=0$. For the
hadronic EoS the extension to finite temperature is well known and
rather straightforward (see e.g. \citet{Glendenning}).  For the quark
EoS, it is trivial to compute finite temperature effects for
non-interacting QM, much less so for matter in which diquark
condensate can take place.  We will therefore discuss finite
temperature effects only in the case of non-interacting quarks.  As it
will be clear in the next Sections, our results should not be too much
affected by this limitation.

\subsection{EoS of hadronic matter}

Concerning the hadronic phase, we use a relativistic self-consistent
theory of nuclear matter in which nucleons interact by exchanging
virtual isoscalar and isovector mesons ($\sigma,\omega,\rho$)
\citep{Glendenning91}.  At $T=0$, in the mean field approximation and
for an infinite and homogeneous system,
the thermodynamic potential per unit volume $\Omega$ can be written as:
\begin{eqnarray}
\Omega=-\frac{1}{3 \pi^2}\sum_{B} \int_0^{k_{FB}}\!\!\!\! {\rm d}k
\frac{k^4}{E_{B}^\star(k)}+\frac{1}{2} m_\sigma^2\sigma^2 
+ \frac{1}{3} a \sigma^3+\frac{1}{4}b\sigma^4
-\frac{1}{2}m_\omega^2\omega^2_0-\frac{1}{2}m_{\rho}^2 \rho^2_{03} \, ,
\end{eqnarray}
where the sum runs over the eight baryon species, 
${E_{B}^\star(k)}=\sqrt{k^2+{M_{B}^\star}^2}$ and 
the baryon effective masses are ${M_{B}^\star}=M_{B}-g_\sigma \sigma$. 
Only the temporal components of the vector fields (and the third
isospin component of $\rho$, $\rho_{03}$) appear at mean field level.
Since we will be interested in computing $\beta$-stable matter, let us
also write the chemical potentials $\mu_B$ which are connected to the
effective chemical potentials $\mu_B^\star$ and to the vector meson
fields through the relation:
\be
{\mu_B^\star}=\mu_B  - g_\omega\omega -t_{3B} g_{\rho}\rho_{03} \,  , 
\ee
where $t_{3B}$ is the isospin 3-component for 
baryon $B$ and the relation to the Fermi momentum $k_{FB}$ is provided by 
$\mu_B^\star=\sqrt{k^2_{FB}+{M_B^\star}^2}$. 
The isoscalar and isovector meson fields ($\sigma$, $\omega$ and
$\rho$) are obtained as a solution of the field equations in mean
field approximation and the related couplings ($g_\sigma$, $g_\omega$
and $g_\rho$) are the parameters of the model
\citep{Glendenning91,Knorren95}.

\begin{figure}[!ht]
\parbox{14cm}{
\scalebox{0.6}{
\includegraphics*[-40,500][560,800]{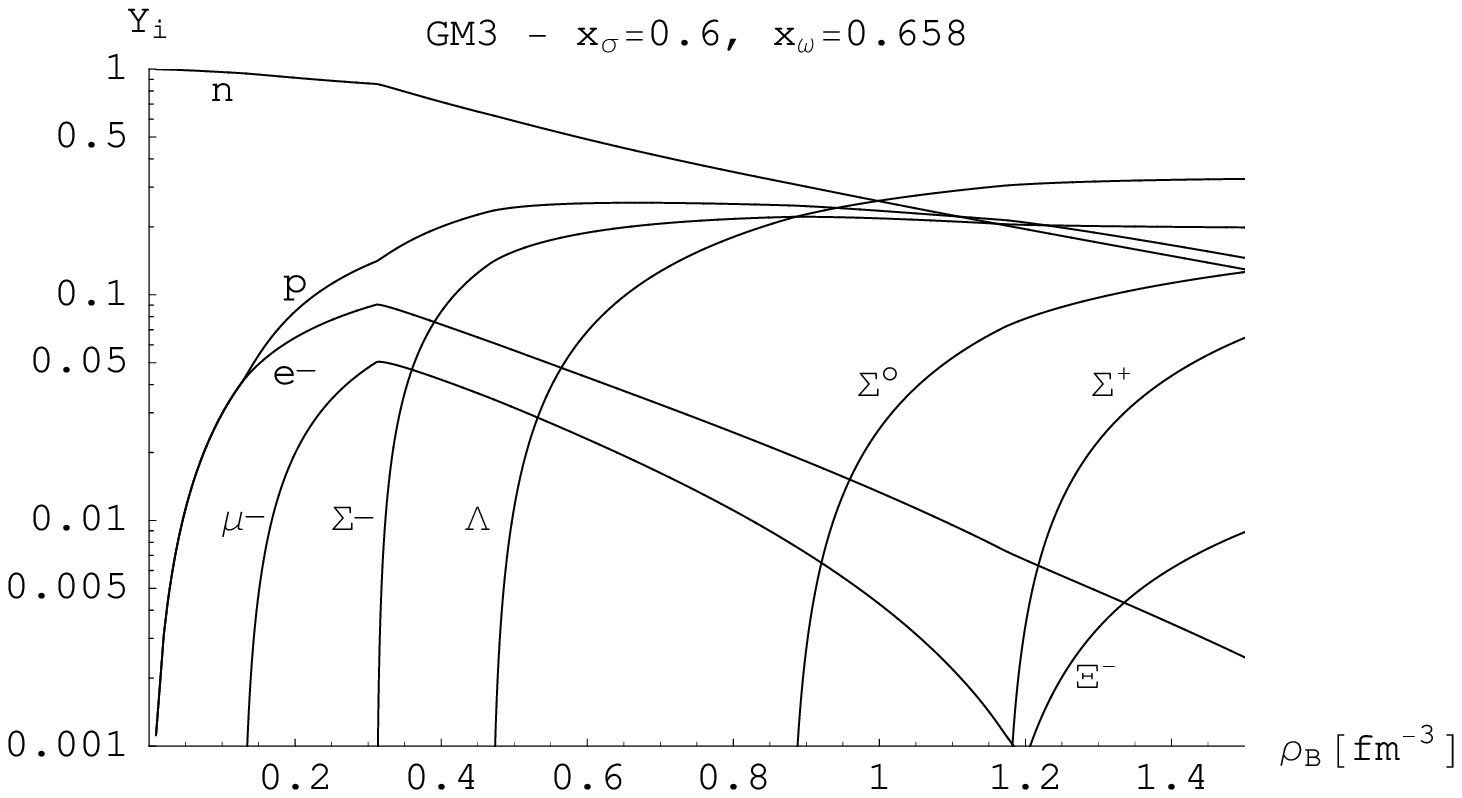}}
\caption{
\footnotesize
Particle fractions $Y_i$ of neutral and $\beta$-stable hadronic matter 
as a function of baryonic density $\rho_B$ for the GM3 
hadronic equation of state of \citet{Glendenning91}.\label{fighadronic}}
}
\end{figure}

\noindent In Fig. \ref{fighadronic} we display the relative
concentrations of the various particle species $Y_i=\rho_i/\rho_B$ as
a function of baryonic density $\rho_B$ by imposing charge neutrality
and $\beta$-equilibrium for the GM3 parameter set
\citep{Glendenning91}.  As it can be seen in the Figure, hyperons
start being produced at rather low densities.  While this is not a
problem in a relativistic mean field model, since stars having a large
enough mass can be obtained, in a non relativistic approach the
softening of the EoS would be too extreme \citep{Baldo:1999rq}. For
this reason, we will discuss both EoSs in which hyperons are taken
into account and also pure nucleonic ones.

\subsection{EoS of quark matter}
\label{doublescenario}

The central density of compact stellar objects may reach values up to
ten times nuclear matter saturation density and therefore it is common
opinion that a phase transition to deconfined QM may take place at
least in the central region.  Recent studies on the QCD phase diagram
at finite densities and temperatures have revealed the existence of
several possible types of quark phases. Many theoretical works have
investigated the possible formation of a diquark condensate in the
quark phase, at densities reachable in the core of a compact star
\citep{alf5,alf1,Alford:2004hz,Ruster:2005ib} and the formation of
this condensate can deeply modify the structure of the star
\citep{alf4,baldo1,Blaschke:2003rg,lugo1,lugo2,Drago:2004vu,Blaschke:2005uj,Lavagno:2005zj}.
It is widely accepted that the Color-Flavor Locking (CFL) phase is the
real ground state of QCD at asymptoticly large densities.  On the
other hand, since the structure of the QCD phase diagram depends
strongly on the value of the strange quark mass $m_s$ and on the
diquark coupling parameters, at this stage the complex structure of
the QCD phase diagram can not be completely determined at finite
densities.  In particular, in addition to a phase of unpaired Normal
Quark matter (NQ) present at low densities, several superconducting
phases (such as 2SC or gapless CFL) can occur at the large baryon
densities reached in the center of a compact star.

\noindent The transition from hadronic to CFL phase could proceed in
two steps, first with a transition from hadronic matter to a 2SC phase
(or to NQ, depending on the model parameters) and then from 2SC to
CFL.  In the scheme proposed in \citet{Drago:2005qb}, the first
transition takes place due to the increase of the baryonic density
(due to mass accretion), while the second transition is associated
with the deleptonization (and the cooling) of the newly formed star
containing 2SC phase.  These two transitions can both be first order
\citep{Ruster:2005ib} and therefore the newly formed hybrid or quark
star containing 2SC quark matter can become metastable and then decay
into a star containing CFL phase with a characteristic time delay
which corresponds to the nucleation time of a drop of CFL phase inside
the 2SC phase.  It is interesting to remark that this scenario is
compatible with an analysis of the time-structure of the light curves
of GRBs \citep{Drago:2005cc}.

\noindent In this paper we consider therefore two different structures
for the QM phase: a NQ phase, described by the MIT bag model and a CFL
phase.  Since we are interested in the bulk properties of a compact
star, for the CFL phase we adopt the simple scheme proposed by
\citet{alf4} where the binding energy $\Delta$ of the diquark
condensate in the thermodynamic potential is expanded up to order
$(\Delta/\mu)^2$ and the gap is assumed to be independent on the
chemical potential $\mu$.
Following this approximation the thermodynamic potential can be written as:
\bq
\Omega_{CFL}=\frac{6}{\pi^2}\int_0^\nu k^2(k-\mu)\,{\mathrm d}k
+
\frac{3}{\pi^2}\int_0^\nu k^2(\sqrt{k^2+m_s^2}-\mu)\,{\mathrm d}k-
\frac{3 \Delta^2 \mu^2}{\pi^2} \; ,
\eq
with
$
\nu=2\mu-\sqrt{\mu^2+\frac{m_s^2}{3}}\, ,
$
and the quark density $\rho$ is calculated numerically
by deriving the thermodynamic potential respect to $\mu$. 
Pressure and energy density read
\bq
P&=&-\Omega_{CFL}(\mu)-B-\Omega^{e}(\mu_e) \; ,\\
E/V&=&\Omega_{CFL}(\mu)+\mu\rho+B+\Omega^{e}(\mu_e)+\mu_e\rho_e \; .
\eq 

\subsection{Mixed phase of hadrons and quarks}
\label{fasemista}
The transition from nuclear matter to QM can proceed via a mixed 
phase if the surface tension at the interface separating quarks and hadrons 
is not too large \citep{Heiselberg:1992dx,Voskresensky:2002hu,Bejger:2005am}.
Gibbs conditions have to be satisfied in the presence of two conserved
charges, the baryonic (B) and the electric (C) one 
(\citet{Glendenning:1992vb}, see also \citet{Drago:2001nq}),
whose conservation laws read:
\bq 
\rho_B&=&(1-\chi)\rho_B^h+\chi\rho_B^q \, ,\\
\rho_C&=&(1-\chi)\rho_C^h+\chi\rho_C^q+\rho_e=0\,\nonumber .
\eq
Here $\chi$ is the fraction of matter in the quark phase and the 
superscripts $h$ and $q$ label the densities in the hadronic and in the
quark phase, respectively. The electron charge density $\rho_e$
contributes to make the total electric charge equal to zero.\\
The EoS appropriate to the description of a compact star
has also to satisfy $\beta$-stability conditions. The equations
of chemical equilibrium under $\beta$-decay and deconfinement reactions 
are the following:
\bq
\mu_n-\mu_p=\mu_e \ \  &,& \quad
\mu_n-\mu_p=\mu_\mu \nonumber \, ,\\
2\mu_d+\mu_u=\mu_n\ \  &,& \quad\mu_u-\mu_d=\mu_p-\mu_n\nonumber \, ,\\
\mu_s&=&\mu_d\,.
\eq
Finally, the usual condition for mechanical equilibrium, {\it i.e.}
the equality of the pressure in the two phases, has to be imposed and it reads:
\be
P^h=P^q \,.
\ee

The previous equations have to be solved together with the
field's equations for the adopted hadronic and quark models.

As already mentioned in the Introduction and as it will be further clarified 
in the next Sections, 
to our purposes it is also important to consider the EoS of a mixed-phase 
in which $\beta$-equilibrium has not yet been reached.
A crucial analysis has been made in a previous paper
\citep{Drago:2003wg} where it has been studied in detail the process of
formation of mixed phase during a perturbation (see in particular
Sec. IIIB). The most important result that
we can borrow from that analysis, is that it is possible to impose
Gibbs conditions, only based on rapid processes, and that through
those conditions a mixed phase, stable on the dynamical timescale, can
be produced.  
Indeed,
if a detonation can take place the conversion process can be so rapid
that $\beta$-equilibrium is reached only after a (short) delay.
This non $\beta$-stable mixed-phase is defined to have, at any density, the 
same isospin ratio $Z/A$ of pure hadronic matter. 

\begin{figure}[!hb]
\parbox{14cm}{
\scalebox{0.6}{
\includegraphics*[-40,510][515,775]{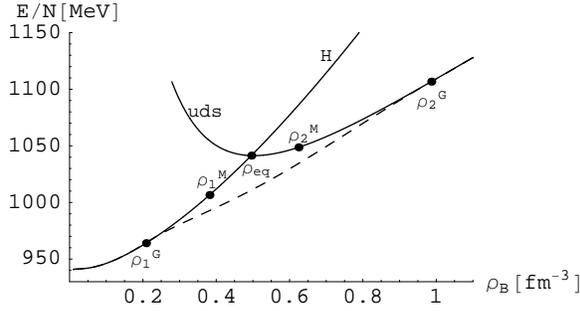}}
\caption{
\footnotesize
Typical scheme of a first order transition at finite density with
mixed phase.  The solid lines represent the pure hadronic phase (H)
and the pure quark phase (uds).  The dashed line, starting at
$\rho_1^G$ and ending at $\rho_2^G$, is the mixed phase obtained
imposing Gibbs conditions and for a vanishing surface tension. When
the surface tension increases the region of mixed phase shrinks and it
reduces to the one obtained using the Maxwell construction which
starts at $\rho_1^M$ and ends at $\rho_2^M$.  $\rho_{eq}$ is the
density at which the energies of the two pure phases are equal.
\label{fig-mixed}}
}
\end{figure}

In Fig.~\ref{fig-mixed} we show an example of mixed phase EoS. Both
the results of the Gibbs and of the Maxwell constructions are
displayed.  The Gibbs construction corresponds to a vanishing surface
tension while the Maxwell construction corresponds to the situation in
which the surface tension is so large that it is no more convenient to
form finite size structures and no mixed phase would exist inside a
compact star.  If the surface tension is finite but smaller than
$\sigma\approx$ 30 MeV/fm$^2$ (as estimated by
\citet{Voskresensky:2002hu}) the mixed phase shrinks respect to the
one obtained using the Gibbs construction and complicated structures
have to develop in order to minimize the energy of the system.
In Fig.~\ref{fig-mixed} we also indicate a density $\rho_{eq}$ such that for 
$\rho>\rho_{eq}$ it is energetically convenient to transform completely 
hadrons into quarks, although the energy of the system can be further reduced  
forming a mixed phase.

\begin{figure}[t]
\parbox{14cm}{
\scalebox{0.6}{
\includegraphics*[-40,500][515,800]{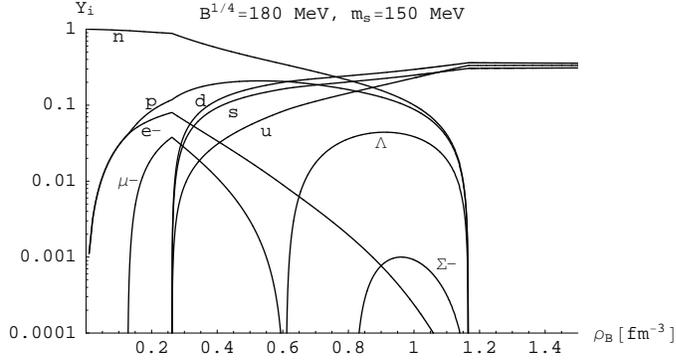}}
\caption{
\footnotesize
Particle fractions $Y_i$ of neutral and $\beta$-stable 
hadronic and quark matter 
as a function of baryonic density $\rho_B$ for the same GM3 
hadronic equation of state used in Fig.~\ref{fighadronic}
and using the MIT bag model with $B^{1/4}=180$ MeV to describe the quark phase.
\label{fighadronicquark}}
}
\end{figure}

\noindent In Fig. \ref{fighadronicquark} we display the relative
concentrations of the various particle species $Y_i=\rho_B^i/\rho_B$ as
a function of baryonic density $\rho_B$ by imposing charge neutrality
and $\beta$-equilibrium for the GM3 parameter set
\citep{Glendenning91} and using $B^{1/4}$ = 180 MeV.  Quarks start
being produced, in the mixed phase, at rather low densities.
Comparing with Fig. \ref{fighadronic}, we notice that hyperons density
is greatly suppressed and the role of hyperons in neutralizing
$\beta$-stable matter is now played mainly by quarks.

\subsubsection{Thermal nucleation of the mixed phase \label{tn}}

For finite values of the surface tension complicated structures
develop in the mixed phase, traditionally called drops, rods and slabs.
In order to understand under which conditions a fluidodynamical
description of the formation of mixed phase is realistic,
we have to estimate the dynamical time-scale of the formation of these
structures. To this purpose we can borrow again from the analysis
of \citet{Drago:2003wg} where two quantities are compared, the size
of the barrier which the system has to overcome in order to form a 
structure and the size of the perturbation of
the system. In \citet{Drago:2003wg} the perturbation was due to a gravitational
instability and it was estimated to be of the order of a few MeV per baryon.
Here the perturbation of the system can be estimated from the size of the
mechanical shock (if the process is a detonation) or from the temperature
(if it is a deflagration). Since we will see that the conversion always
proceeds as a deflagration, we can compare the temperature reached by the
system immediately after the conversion with the height of the barrier.
If the temperature is not much lower than the height of the barrier the 
structure formation proceeds thermally and it is very rapid 
\citep{madsen1,madsen2,DiToro:2006pq}. 
Instead, new structures
can form only via quantum nucleation, typically a very slow process.

\begin{figure}[b]
\begin{center}
\includegraphics*[scale=0.4]{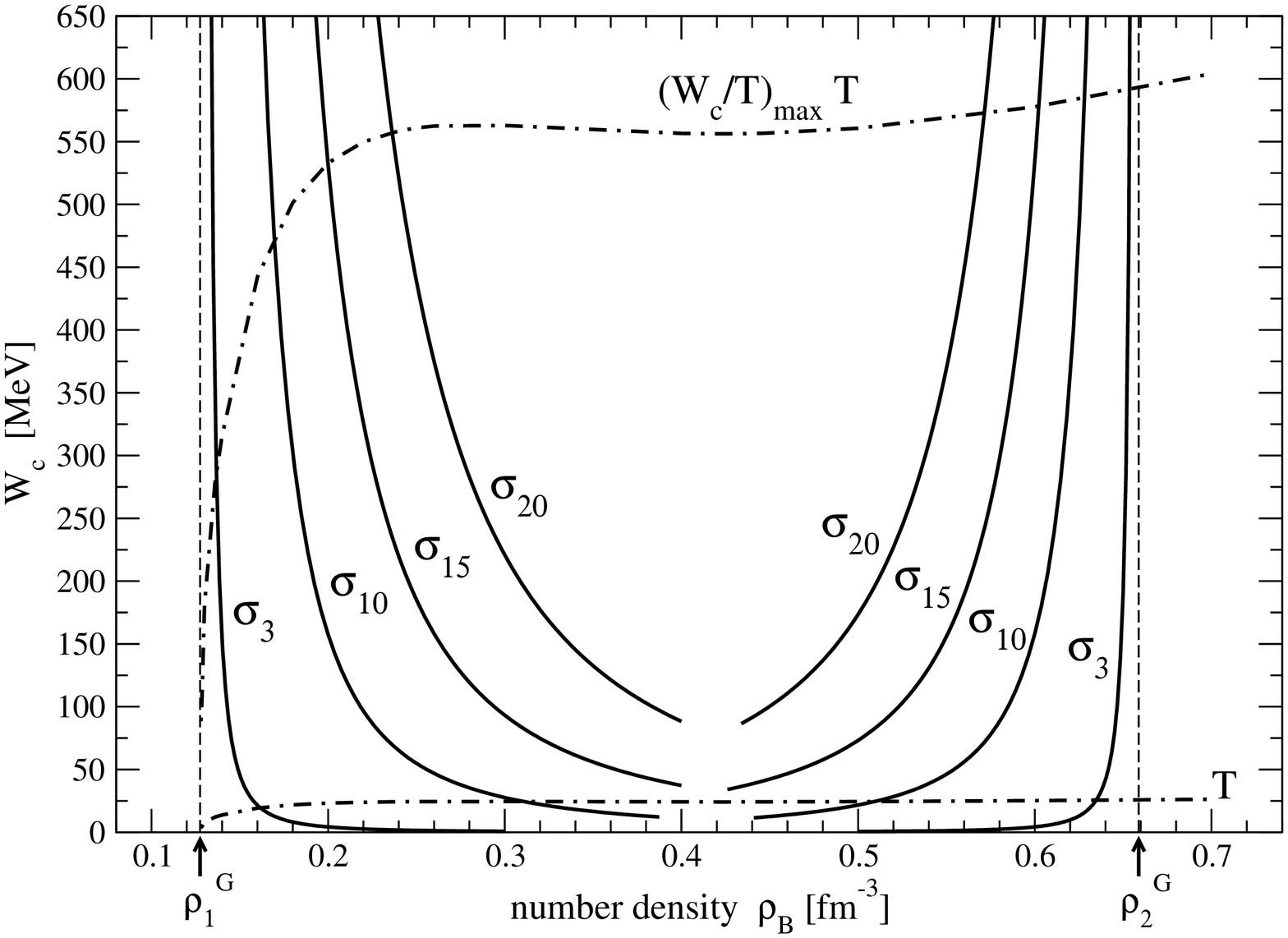}
\end{center}
\parbox{14cm}{
\caption{
\footnotesize 
Work needed to form a bubble of new phase as a
function of baryonic density $\rho_B$ (solid lines). 
The four lines at lower density correspond to the formation of bubbles
of quarks in the hadronic medium (for four different values
of the surface tension $\sigma$). The four lines
at higher density correspond to the formation of bubbles of
hadrons in a pure quark matter medium.
Here we have used the GM3 EoS to describe the hadronic phase and 
the MIT bag model with $B^{1/4}=165$ MeV to describe the quark phase. 
$\beta$-stability has been imposed in both phases.
The starting and the ending point of the mixed phase using the Gibbs 
construction are shown ($\rho_1^G$ and $\rho_2^G$).
We also show the temperature T reached by the system due to the 
exothermic deconfinement process and the product between T and 
$(W_c/T)_{\mathrm{max}}$, indicating the maximum value of the work
for which thermal nucleation can take place and be rapid.
\label{fig-barriera}}
}
\end{figure}

The thermal nucleation rate of drops can be estimated as 
\citep{madsen1,madsen2}:
\be
{\mathcal R}=\mu^4 \exp (-W_c/T)\label{rate}
\ee
where $\mu$ is the baryon chemical potential and $W_c=W(R_c)$ 
represents the work needed to form the smallest bubble capable of growing.  
$W_c$ corresponds to the maximum of the free energy of the 
bubble of the new phase. The free energy reads:
\be
W(R)=-\frac{4\pi}{3}R^3 [(P_2-P_1) - \rho_2 (\mu_2-\mu_1)]+4 \pi\sigma R^2
\ee
and its maximum as a function of $R$ (which has to be computed at $\mu_2=\mu_1$
like in \citet{madsen2}) is:
\be
W_c=\frac{16\pi}{3}\frac{\sigma^3}{(P_2-P_1)^2}
\ee
which is obtained for a radius $R=R_c=2\sigma/(P_2-P_1)$.
In Eq.~(\ref{rate}) the prefactor has been
taken as the chemical potential, while, when thermal nucleation 
is investigated at very large temperatures, 
as the ones reached in ultrarelativistic heavy ion scattering,
the prefactor is related to the temperature.
The number of bubbles of new phase formed inside the old
phase\footnote{In the low density region of the mixed phase
there are drops of quarks in a medium of hadrons. In the upper
density region the situation is reversed. We can still apply
the formalism here introduced if we suppose that first pure quark matter
forms (where energetically convenient) and immediately after
drops of hadrons are produced in the quark matter phase.}
in a volume $V$ and in a time $t$ is given by: 
\be
{\mathcal N}={\mathcal R}\,V\, t\,\, , 
\ee
Let us call $\lambda$ the spacing between two drops in the mixed phase.
The number of drops in a volume $V$ is given by $V/\lambda^3$.
{\it{A fluidodynamical description of the formation of mixed phase is 
realistic if the number of bubbles produced while the front moves 
over a distance $\lambda$ is of the order of the number of bubbles that 
have to be present in the mixed phase}}:
\be
{\mathcal R}\,V\,t={\mathcal R}\,S\,\lambda\left(\frac{\lambda}{v}\right)
\ge\frac{V}{\lambda^3}=\frac{S}{\lambda^2}.
\ee
Here $v$ is the velocity of the front and $S$ is the area of the surface 
of the front.
Therefore the following constraint has to be satisfied:
\be
\frac{W_c}{T} \, \, \le \, \, \text{ln} \left( \frac{\mu^4\,\lambda^4}{v} 
\right) \equiv 
\left( \frac{W_c}{T}\right)_{\text{max}}.
\ee
For instance, in the case of $\beta$-equilibrium mixed phase and $B^{1/4}$ = 
165 MeV we obtain $(W_c/T)_{\text{max}}=23$.
For the same parameter in Fig.~\ref{fig-barriera} we display 
$W_c$ as a function of the baryonic density $\rho_B$ 
(for different values of the surface tension $\sigma$). 
In the same figure we also show the temperature T (calculated as 
described in Sec.~\ref{temperature}) and the product between T and 
$(W_c/T)_{\text{max}}$. From this figure it is 
possible to understand in which density regions the fluidodynamical 
description is allowed.
In the density regions where $W_c< T \cdot (W_c/T)_{\text{max}}$ 
the mixed phase forms without delay, the fluidodynamical description of 
the transition is correct and matter goes directly 
into the mixed phase. Instead, where $W_c> T \cdot (W_c/T)_{\text{max}}$ 
 the formation of the 
mixed phase is delayed. In this latter case matter, immediately after the 
front, can not be in the mixed phase. 
As an example, let us discuss the case $\sigma=10$ MeV/fm$^2$ shown in 
Fig.~\ref{fig-barriera}. We can identify three regions. In the central 
region, with $\rho_B$ between 0.17 fm$^{-3}$ and 0.628 fm$^{-3}$, 
 $W_c< T \cdot (W_c/T)_{\text{max}}$
 and the fluidodynamical picture is always allowed.
In the region between 0.628 fm$^{-3}$ and $\rho_2^G$ it is energetically 
convenient to transform completely hadrons into quarks (although the 
energy of the system can be further reduced by forming a mixed phase) and the 
fluidodynamical picture can still be applied by burning completely hadrons 
into quark. Finally, in the region between 
$\rho_1^G$ and 0.17 fm$^{-3}$ it is neither possible to form mixed phase soon 
after the front nor it is convenient to form pure quark phase 
(because $\rho_B$ is less then $\rho_{\text{eq}}$). 
Therefore in that region the process of formation of mixed phase can not be 
described fluidodynamically and it can be rather slow.

\noindent It is important to note that this description of thermal 
nucleation is related to the formation of drops of new phase. 
If the surface tension $\sigma$ is too large  
the region where 
the formation of mixed phase is allowed becomes the region in which we 
have to form more complicated structures to minimize the energy, like 
rods or slabs. In this case our simplified picture is not applicable.

\noindent In conclusion, if the surface tension vanishes the 
fluidodynamical description of the transition is realistic down to $\rho_1^G$.
If the surface tension is finite but small (until few ten MeV/fm$^2$) it is 
realistic but only down to a density larger then $\rho_1^G$. Finally, if the 
surface tension is large the fluidodynamical picture can be applied down to 
$\rho_{\text{eq}}$, but assuming matter to first transform (during the 
hydrodynamical timescale) to a pure quark matter phase.

The process of absorption of an hadron into the pure quark matter phase 
could also be described phenomenologically as the fusion of a small bubble 
of quarks (the hadron) into a much larger bubble. In the density region in 
which pure quark matter can form the large energy gain should make negligible 
the role played by surface tension, but for very large values of the latter.

\subsection{Compact star structure}

The EoSs analyzed in the last subsections can be used
to compute the structure of compact stars 
by solving the Tolman-Oppenheimer-Volkoff system
of equations.

\begin{figure}[!b]
\parbox{14cm}{
\scalebox{0.65}{
\includegraphics*[-40,510][515,775]{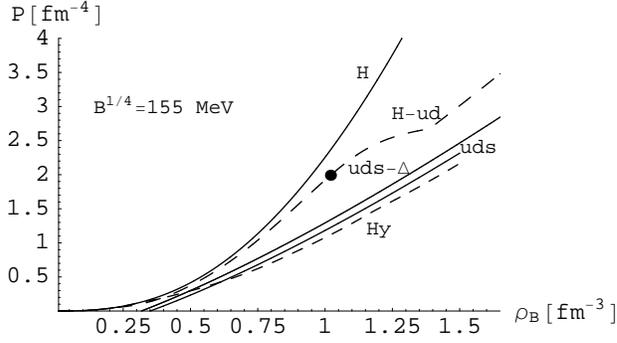}}
\caption{
\footnotesize 
Pressure as a function of the baryon density for pure nucleonic phase
(H), not $\beta$-stable mixed phase of nucleonic matter and unpaired
quark ud matter (H-ud), unpaired uds quark matter (uds), pure CFL
phase (uds-$\Delta$).  The dot on the not $\beta$-stable EoS
corresponds to $\rho_{eq}$, defined in Fig.~\ref{fig-mixed}.  Here
$B^{1/4}$ = 155 MeV.
\label{fig-eos155}}
}
\end{figure}

\begin{figure}[ht]
\parbox{14cm}{
\scalebox{0.65}{
\includegraphics*[-40,510][515,775]{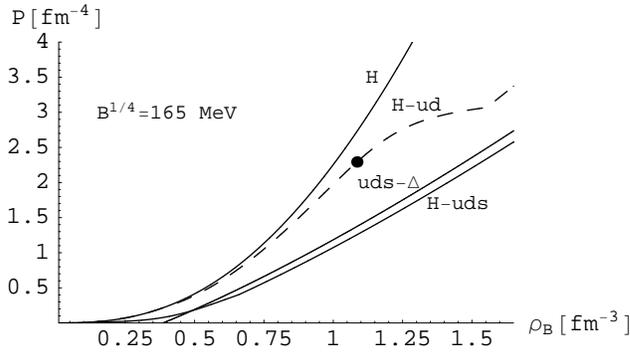}}
\caption{
\footnotesize 
Pressure as a function of the baryon density for pure nucleonic phase
(H), not $\beta$-stable mixed phase of nucleonic matter and unpaired
quark ud matter (H-ud), $\beta$-stable (mixed) phase of hadronic and
unpaired uds quark matter (H-uds), pure CFL phase (uds-$\Delta$).  The
dot on the not $\beta$-stable EoS corresponds to $\rho_{eq}$, defined
in Fig.~\ref{fig-mixed}.  Here $B^{1/4}$ = 165 MeV.
\label{fig-eos165}}
}
\end{figure}

\begin{figure}[hb]
\parbox{14cm}{
\scalebox{0.65}{
\includegraphics*[-40,510][515,775]{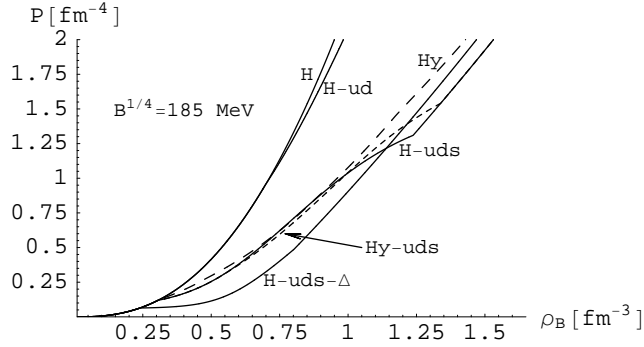}}
\caption{
\footnotesize 
Pressure as a function of the baryon density for pure nucleonic phase
(H), hyperonic matter (Hy), not $\beta$-stable mixed phase of
nucleonic matter and unpaired quark ud matter (H-ud). Also displayed
are $\beta$-stable (mixed) phases made of nucleons and unpaired uds
quarks (H-uds), of hyperons and unpaired uds quarks (Hy-uds) and,
finally, of nucleons and CFL quarks (H-uds-$\Delta$).  Here $B^{1/4}$
= 185 MeV.
\label{fig-eos185}}
}
\end{figure}

In Figs.~\ref{fig-eos155},\ref{fig-eos165},\ref{fig-eos185} we show
the pressure as a function of the baryon density for the various EoSs
discussed in the previous subsections.  For the pure hadronic phase we
adopt the GM3 parameters of \citet{Glendenning91}, either using only
nucleonic degrees of freedom (label H) or including hyperons (label
Hy).  Concerning the pure quark phase, we consider both NQ in
$\beta$-equilibrium (label uds) and a CFL phase (uds-$\Delta$) in
which a constant value of the gap has been used ($\Delta=100$ MeV).
In Fig.~\ref{fig-eos155} the value of the pressure of the vacuum is
$B^{1/4}$=155 MeV.  If the system is allowed to reach
$\beta$-stability, mixed phases cannot be formed for such a low value
of $B$. Moreover hyperons are not present in the $\beta$-stable EoSs
because
they are completely substituted by quarks.
On the other hand, if the system has not yet reached $\beta$-stability, 
mixed phase can exist and in Fig.~\ref{fig-eos155} we show an example
of such a phase (label H-ud) formed from nucleonic matter and unpaired quarks.
In such a case, only up and down quarks can be present since
strangeness is absent in the hadronic phase and we assume that
there is no time for weak interactions to take place. 
In Fig.~\ref{fig-eos165} we assume $B^{1/4}$=165 MeV.
Here a mixed phase can be formed at moderate densities if the quarks
are not gapped, but again we obtain pure QM if the formation of a
strongly gapped diquark condensate is taken into account.  Finally, in
Fig.~\ref{fig-eos185} we use $B^{1/4}$=185 MeV.  At variance with
Figs.~\ref{fig-eos155},\ref{fig-eos165} here the hyperons can be
present together with quarks, at least if the latter do not form a
condensate.  We will not discuss even larger values of $B$ since, as
it will be clear in the following, $B^{1/4}$=185 MeV is in our scheme
the largest value of the pressure of the vacuum for which a rapid
formation of a mixed phase of deconfined quarks can take place, if a
small, but non vanishing surface tension at the interface between
hadrons and quarks is taken into account.

\begin{figure}[ht]
\parbox{14cm}{
\scalebox{0.7}{
\includegraphics*[-40,510][515,775]{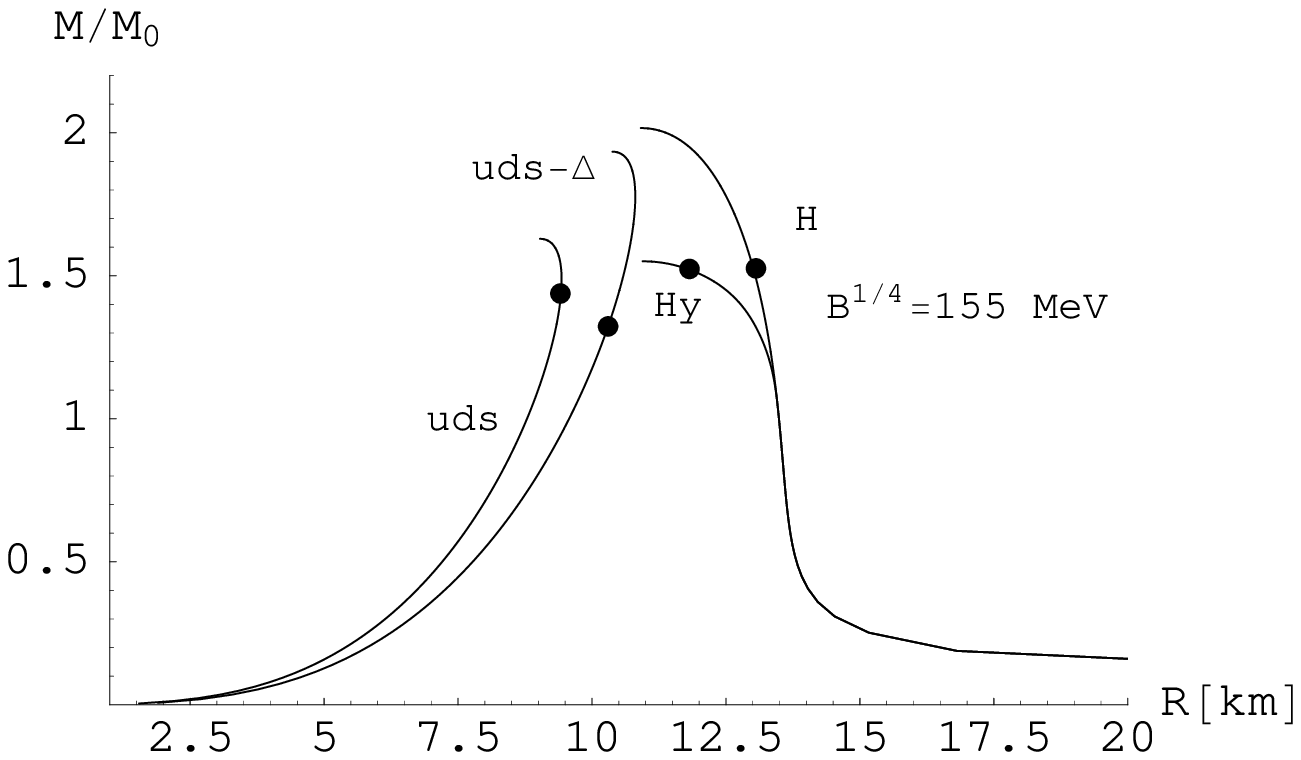}}
\caption{
\footnotesize 
Mass-radius relations of nucleonic stars (H), hyperonic stars (Hy),
quark stars made of unpaired quark matter (uds) and of color
superconducting CFL phase (uds-$\Delta$). The dots indicate stars
whose baryonic mass is $M_B$ = 1.7 $M_\odot$.  Here $B^{1/4}$ = 155
MeV.
\label{fig-mr155}}
}
\end{figure}

\begin{figure}[hb]
\parbox{14cm}{
\scalebox{0.7}{
\includegraphics*[-40,510][515,775]{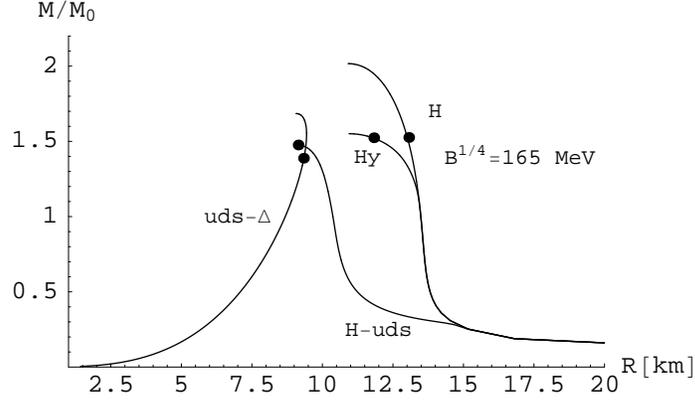}}
\caption{
\footnotesize 
Mass-radius relations of nucleonic stars (H), hyperonic stars (Hy), 
hybrid stars made of hadrons and of unpaired quark matter (H-uds) 
and quark stars made of color superconducting CFL phase (uds-$\Delta$). 
The dots indicate stars whose baryonic mass
is $M_B$ = 1.7 $M_\odot$. Here $B^{1/4}$ = 165 MeV.
\label{fig-mr165}}
}
\end{figure}

\begin{figure}[ht]
\parbox{14cm}{
\scalebox{0.7}{
\includegraphics*[-40,510][515,775]{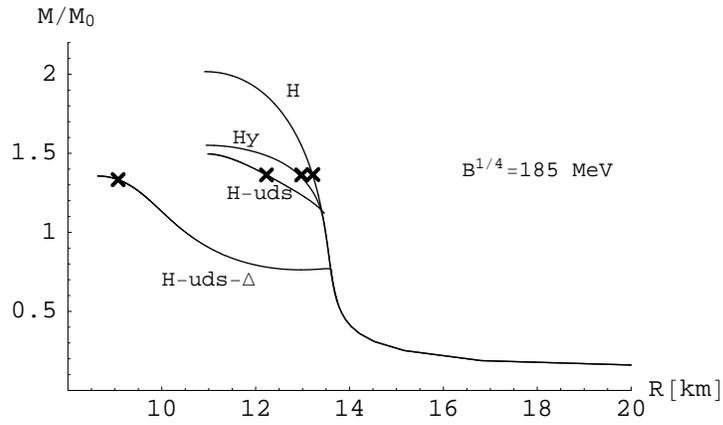}}
\caption{
\footnotesize 
Mass-radius relations of nucleonic stars (H), hyperonic stars (Hy),
hybrid stars made of hadrons and of unpaired quark matter (H-uds) and
of hadrons and CFL quarks (H-uds-$\Delta$). The crosses indicate stars
whose baryonic mass is $M_B$ = 1.5 $M_\odot$. Here $B^{1/4}$ = 185
MeV.
\label{fig-mr185}}
}
\end{figure}

In Figs.~\ref{fig-mr155},\ref{fig-mr165},\ref{fig-mr185} we display the 
mass-radius relations for the same EoSs discussed above.
In Figs.~\ref{fig-mr155},\ref{fig-mr165} we have indicated with
dots stars having a same baryonic mass $M_B = 1.7 M_\odot$.
The gravitational mass of a purely nucleonic star described by the 
GM3 model is then $M_G = 1.53 M_\odot$, and for an hyperonic star
$M_G = 1.52 M_\odot$.
In Fig.~\ref{fig-mr155}, using $B^{1/4}$=155 MeV we obtain
a quark star made of unpaired quarks having $M_G = 1.43 M_\odot$
and a quark star made of CFL quarks whose mass is $M_G = 1.32 M_\odot$.
Similarly, in Fig.~\ref{fig-mr165}, using $B^{1/4}$=165 MeV we obtain
a hybrid star made of nucleons and NQ with a mass
$M_G = 1.48 M_\odot$ and a quark star made of CFL quarks with 
mass $M_G = 1.39 M_\odot$.
Finally, in Fig.~\ref{fig-mr185}, using $B^{1/4}$=185 MeV we obtain
only hybrid stars, whose maximum baryonic mass exceeds only
slightly $M_B = 1.5 M_\odot$ and we indicate with crosses the
stars having that same mass $M_B$. We have not displayed the mass-radius
line for a star made of a mixed phase of hyperons and ungapped quarks,
since it is essentially indistinguishable from the H-uds line.
The reason is that also in this case, although hyperons are present
in the EoS their contribution to the pressure is very small.
Here, fixing $M_B = 1.5 M_\odot$, we obtain $M_G = 1.37 M_\odot$
for the nucleonic star, $M_G = 1.36 M_\odot$ for the hyperonic star,
$M_G = 1.36 M_\odot$ for the hybrid star made of ungapped quarks and,
finally, $M_G = 1.33 M_\odot$ for the hybrid star made of CFL quarks.

\begin{figure}[hb]
\parbox{14cm}{
\scalebox{0.7}{
\includegraphics*[-40,510][515,775]{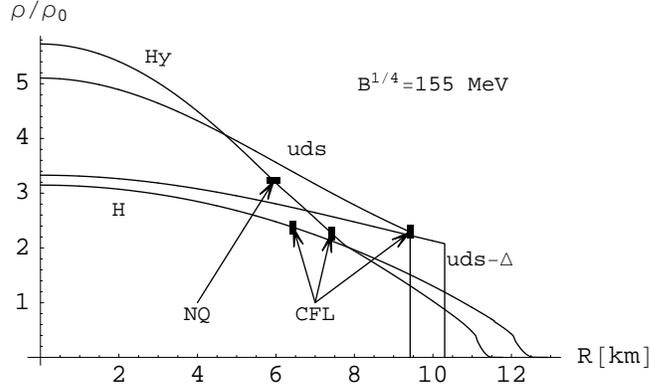}}
\caption{
\footnotesize 
Baryon density profiles for different compact stars having the same
baryonic mass $M_B=1.7 M_\odot$.  Here $B^{1/4}$ = 155 MeV. Labels as
in previous figures.  The arrows indicate the endpoints of the
convective layers (see Sec.~\ref{realistic-convection}).
\label{fig-pro155}}
}
\end{figure}

\begin{figure}[ht]
\parbox{14cm}{
\scalebox{0.7}{
\includegraphics*[-40,510][515,775]{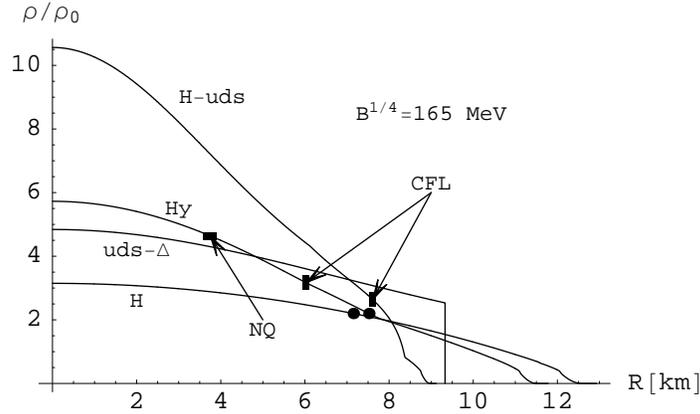}}
\caption{
\footnotesize 
Baryon density profiles for different compact stars having the same
baryonic mass $M_B=1.7 M_\odot$. Here $B^{1/4}$ = 165 MeV. Labels as
in previous figures.  The dots indicate $\rho_{eq}$ for a transition
from nucleonic or hyperonic matter to unpaired quark matter.  When the
transition is to CFL quark matter, a quark star is obtained and the
hydrodynamical argument can be applied to all regions of the star.
The arrows indicate the endpoints of the convective layers (see
Sec.~\ref{realistic-convection}).
\label{fig-pro165}}
}
\end{figure}

\begin{figure}[hb]
\parbox{14cm}{
\scalebox{0.7}{
\includegraphics*[-40,510][515,775]{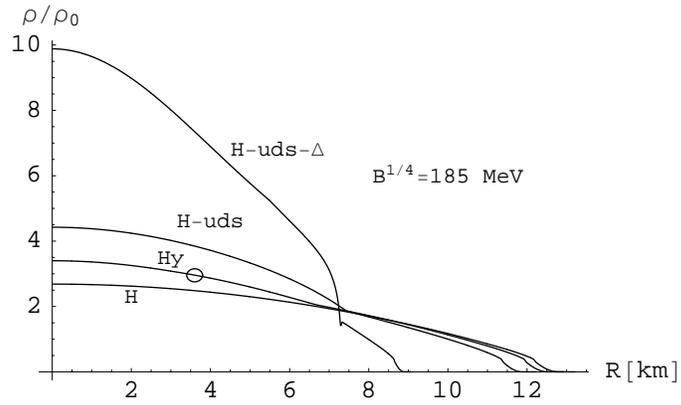}}
\caption{
\footnotesize 
Baryon density profiles for different compact stars having the same
baryonic mass $M_B=1.5 M_\odot$. Here $B^{1/4}$ = 185 MeV. Labels as
in previous figures.  The open circle indicate $\rho_{eq}$ for a
transition from hyperonic matter to CFL quark matter. In all other
cases the central density of the hadronic star was lower than
$\rho_{eq}$.
\label{fig-pro185}}
}
\end{figure}

In Figs.~\ref{fig-pro155},\ref{fig-pro165},\ref{fig-pro185} we show
the density profiles, computed using the same EoSs and for the three
values of $B$.  In Figs.~\ref{fig-pro155},\ref{fig-pro165} the
profiles correspond to a star having $M_B = 1.7 M_\odot$, while in
Fig.~\ref{fig-pro185} the star has $M_B = 1.5 M_\odot$.  In
Fig.~\ref{fig-pro165} we have indicated with two full dots the
position inside the nucleonic and the hyperonic star for which
$\rho_B=\rho_{eq}$, as defined in Sec.~\ref{fasemista}. In
Fig.~\ref{fig-pro155} no dot appears because only quark stars are
formed and therefore at all densities it is energetically convenient
to transform entirely hadrons into quarks. In Fig.~\ref{fig-pro185}
only one open circle appear, in correspondence with the transition
from an hyperonic star to a star containing CFL quarks. In all other
cases, $\rho_{eq}$ is larger than the maximum density of the
corresponding purely hadronic star, indicating that the
fluidodynamical description of the transition cannot be applied if the
surface tension is not low enough, as discussed in~Sec.~\ref{tn}.

\section{Temperature of the quark phase}
\label{temperature}

While we assume in our analysis that the hadronic phase is at $T_H=0$,
the quark or mixed phase produced by the combustion process should be
computed at finite temperature.  The reason is that we are interested
in exothermic processes for which the variation of internal energy is
mainly transformed into heat, while a relatively small fraction of the
internal energy goes into kinetic energy of the newly formed phase.
In the case of an infinite and homogeneous system, which is obviously
a rather poor representation of a star, and in the absence of
dissipative processes, the new phase would continue flowing at a
finite velocity.  In a real star the new phase cannot continue moving
at finite velocity towards the center of the star and some extra heat
is produced by dissipative processes while the kinetic energy
dissipates. The numerical value of the kinetic energy turns out to be
rather small when compared to the heat directly released by the
deconfinement transition (a correction of the order of a few percent in
 the not-$\beta$-stable case)
and therefore we will neglect this contribution.

An estimate of the temperature $T$ of the new phase can be obtained
from thermodynamics first principle.  Since the variation of the
energy per particle $\Delta(E/A)$ depends also on the temperature of
the new phase, the following self-consistent equation has to be
solved:
\be
\Delta \left( \frac{E}{A} \right) (T,\rho_B^h) \equiv 
\frac{e_h(u_h,\rho_B^h,T_h)}{\rho_B^h(u_h)}- 
        \frac{e_q(u_q,\rho_B^q,T)}{\rho_B^q(u_q)} = c_V^q (T-T_h) \, .
\label{termodinamica}
\ee
Here the energy density $e(u,\rho_B,T)$ and the baryon density
$\rho_B(u)$ are computed at finite
velocity $u$ and they read \citep{tolman}:
\bq
& & e(u) = \left( e+p \,u^2\right ) \gamma(u)^2\\
& & \rho_B(u)=\rho_B \, \gamma(u)^2
\eq
where $e$ and $\rho_B$ are the rest frame quantities.
An exothermic process corresponds to a positive $\Delta(E/A)$ and
therefore implies $T>T_h$.

Both sides of Eq.~(\ref{termodinamica}) transform as energies and the
equation is therefore covariant.  It is possible to recast
Eq.~(\ref{termodinamica}) in a form showing more explicitly its
relation with thermodynamics first principle:
\be
\left[ \frac{e_h(u_h,\rho_B^h,T_h)}{\rho_B^h(u_h)}-
\frac{e_q(u_q=u_h,\rho_B^q,T)}{\rho_B^q(u_q=u_h)}\right] -
\left[ \frac{e_q(u_q,\rho_B^q,T)}{\rho_B^q(u_q)}-
\frac{e_q(u_q=u_h,\rho_B^q,T)}{\rho_B^q(u_q=u_h)}\right] = 
c_V^q (T-T_h) \, .
\label{termodinamica2}
\ee
Here the first term in square brackets is the variation of the
internal energy of the system while the second term is the work done
by the system. It is important to stress that all our calculations
are done keeping the volume constant. Due to this assumption, 
the only mechanical work is the one
associated with the kinetic energy of the new phase.
Moreover, no chemical contribution need to be
present, because chemical equilibrium is always assumed, due to the
decoupling of the time-scales and therefore the stoichiometric equation
can be used.

\begin{figure}[ht]
\centering
\parbox{14cm}{
\scalebox{0.6}{
\includegraphics*[14,14][590,750]{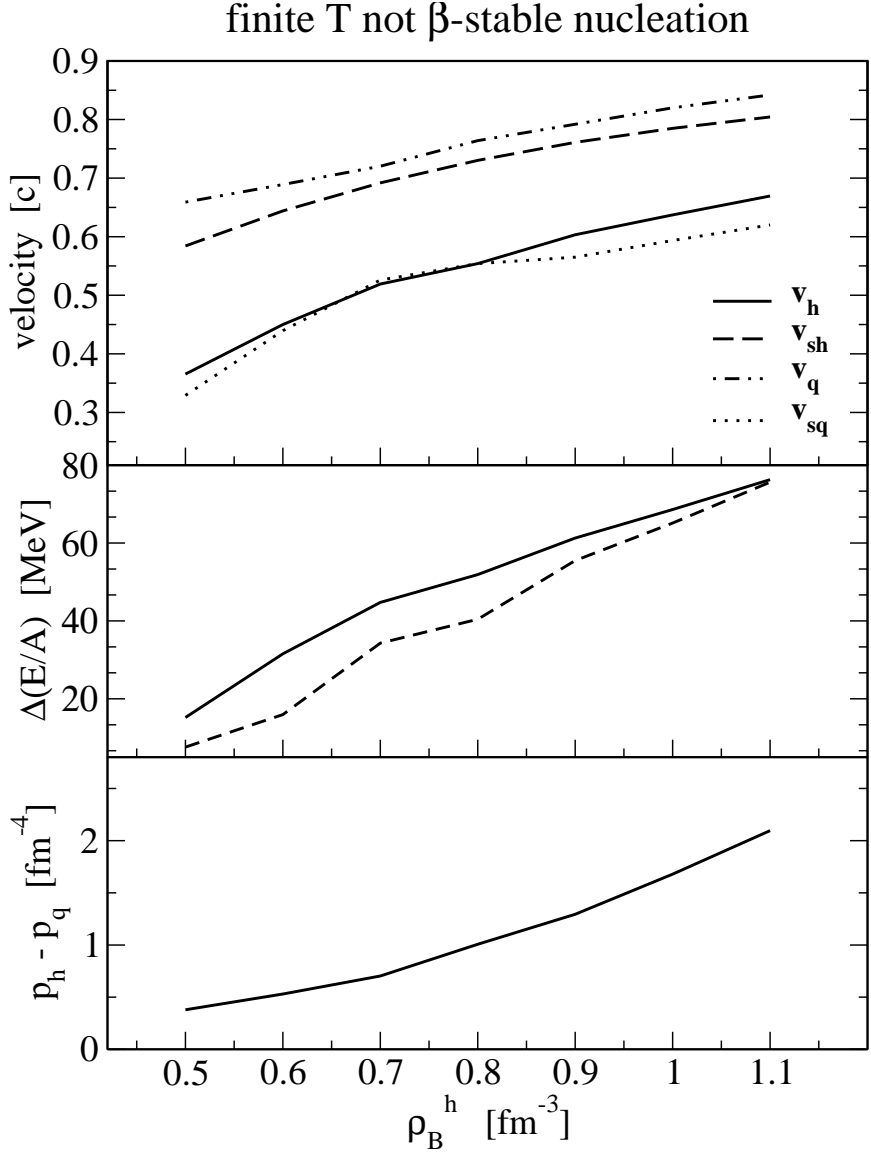}}
\caption{
\footnotesize 
Upper panel: velocity of hadronic phase $v_h$, of the burned phase
$v_q$ and corresponding sound velocities $v_{sh}$ and $v_{sq}$, all in
units of the velocity of light and in the front frame.  Center panel:
energy difference between the two phases (in the hadron phase rest
frame).  The dashed and the solid lines correspond to the first and to
the second iteration in the solution of
Eqs.~(\ref{ene},\ref{mom},\ref{baryon},\ref{termodinamica}).  Lower
panel: pressure difference between the uncombusted and the combusted
phase.  Here the combusted phase is obtained using $B^{1/4}=170$ MeV,
temperatures from 5 to 40 MeV (as estimated from the solid line in the
central panel) and it is not $\beta$-stable.
\label{graf-nb-Tfinita}}
}
\end{figure}

Consistently with the astrophysical scenario discussed in the
Introduction, we will assume that the initial temperature of the
hadronic phase $T_h$ is negligible respect to the final temperature of
the quark (or mixed) phase, i.e.  $T_h\simeq0$. It can be useful to
solve Eq.~(\ref{termodinamica}) in the hadron matter rest frame
($u_h=0$) where quarks flow at a velocity $u_q$ which can be obtained
from Eqs.~(\ref{vh},\ref{vq}).  The solution of
Eqs.~(\ref{ene},\ref{mom},\ref{baryon},\ref{termodinamica}) can be
obtained using an iterative procedure in which the quark phase is
initially assumed at zero temperature when solving
Eqs.~(\ref{ene},\ref{mom},\ref{baryon}) and the new estimate of $T$ is
obtained from Eq.~(\ref{termodinamica}).  In central panel of
Fig.~\ref{graf-nb-Tfinita} we will show the difference between the
first and the second iteration, indicating that convergence can be
reached.

Finally, let us remark that to describe the flow of the new phase in a
real star it would be necessary to solve the hydrodynamical equations
describing the evolution of the system in which a combustion front
propagates \citep{tokareva}. These are rather complicated partial
differential equations which, moreover, should be studied together
with the equations describing the dynamical readjustment of the
star. In our paper we only give a rough estimate of the temperature of
the new phase. As it will be shown in the next Sections, a finite
value of the temperature affects only marginally the mode of
combustion.
 
The numerical value of the temperature was estimated by
\citet{Lugones97} in a different scheme, obtaining results similar to
ours.

\section{Results: deflagration regime \label{deflagration}}

In this Section we discuss the results obtained studying the
fluidodynamics Eqs.~(\ref{ene},\ref{mom},\ref{baryon}) with the models
discussed in Section \ref{eos}.  We are interested in answering two
questions, namely if a detonation is possible and, if this is not the
case, which type of deflagration is obtained.  As already mentioned in
the Introduction, when discussing the possibility of a detonation it
is interesting to discuss also the possibility that matter after the
conversion front is not yet $\beta$-stable, since weak processes are
slower than the strong reactions taking place during deconfinement.
Moreover, the conditions for detonation, Eq.~(\ref{strongdet}), are in
principle more easy to satisfy if matter after the conversion front is
not yet $\beta$-stable, since in that case its EoS is stiffer (as it
can be seen from
Figs.~\ref{fig-eos155},\ref{fig-eos165},\ref{fig-eos185}) and therefore
a larger pressure for the new phase can be obtained.

\begin{figure}[ht]
\centering
\parbox{14cm}{
\scalebox{0.6}{
\includegraphics*[14,14][590,750]{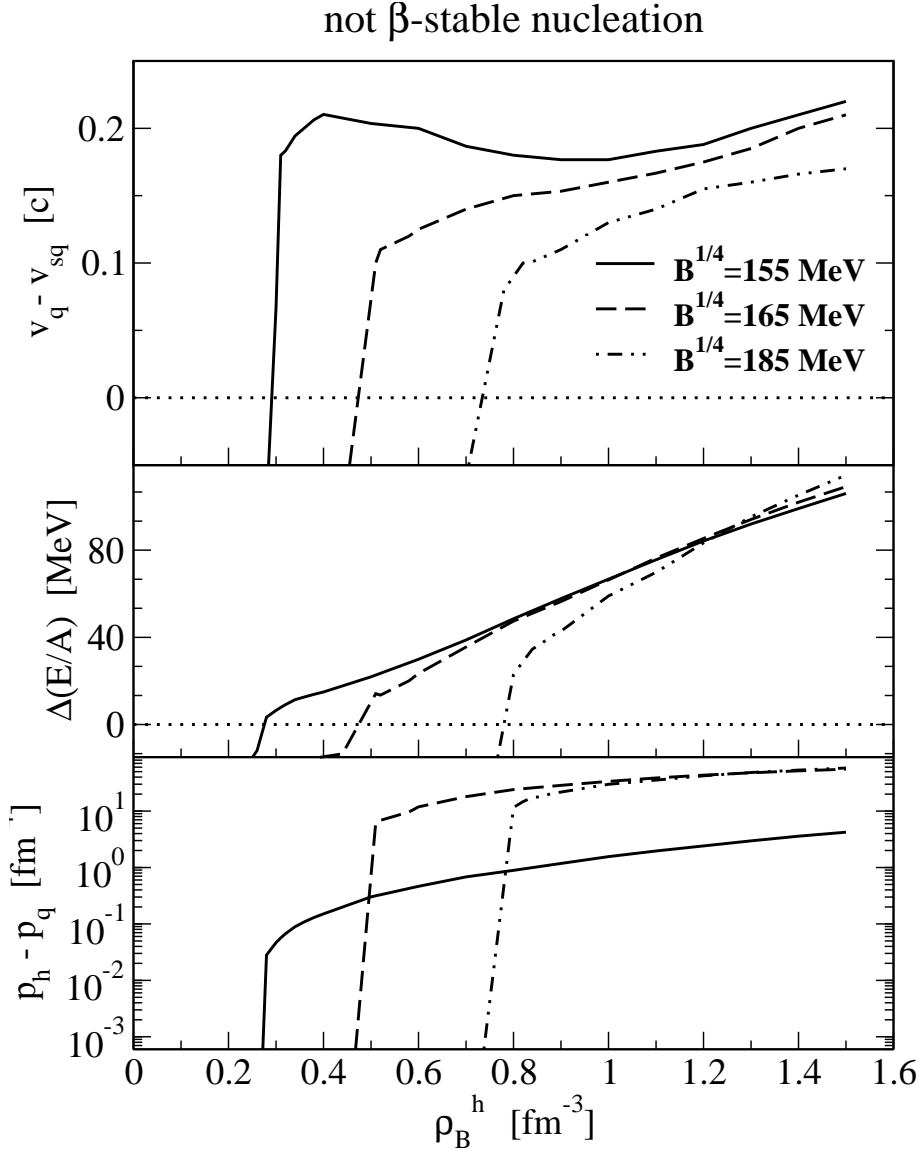}}
\caption{
\footnotesize 
Upper panel: difference between the velocity $v_q$ of the combusted
phase and the sound velocity $v_{sq}$ in the same phase. All the
velocities are in units of the velocity of light and in the front
frame.  The three lines corresponds to different values for the bag
constant.  Center panel: energy difference between the two phases (in
the hadron phase rest frame), as defined in Eq.~(\ref{termodinamica}).
Lower panel: pressure difference between the uncombusted and the
combusted phase.  Here the combusted phase is not $\beta$-stable.
\label{graf-nb}}
}
\end{figure}

\begin{figure}[ht]
\centering
\parbox{14cm}{
\scalebox{0.6}{
\includegraphics*[14,14][590,750]{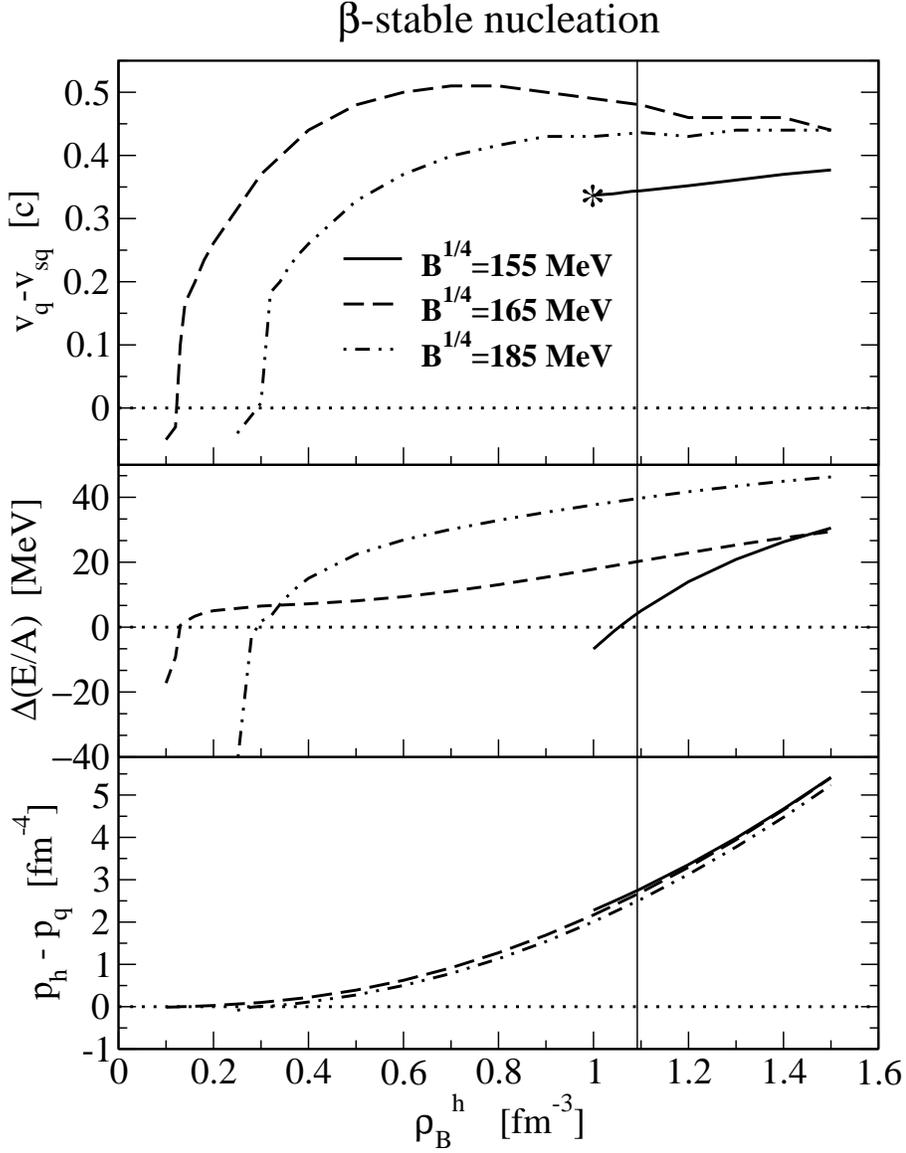}}
\caption{
\footnotesize 
Same as in Fig.~\ref{graf-nb}.  Here the combusted phase is
$\beta$-stable. The star in the upper panel indicates the density
below which the pressure of the newly formed quark matter is negative.
The vertical line corresponds to the central density of the most
massive stable configuration of a nucleonic star obtained using GM3
model ($\rho_h^{max}$=1.09 fm$^{-3}$).
\label{graf-beta}}
}
\end{figure}

\begin{figure}[ht]
\centering
\parbox{14cm}{
\scalebox{0.6}{
\includegraphics*[14,14][590,750]{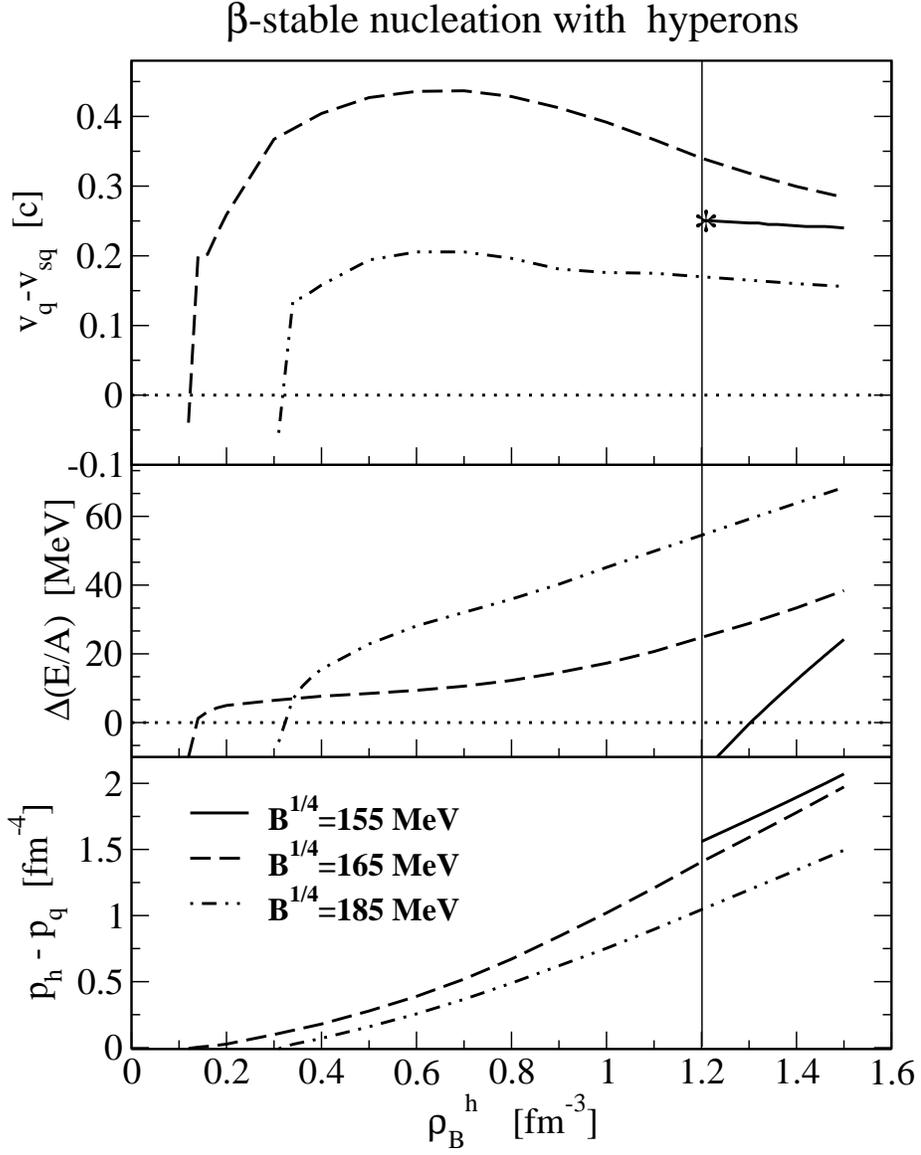}}
\caption{
\footnotesize 
Same as in Fig. \ref{graf-beta}.  Here hyperons are included in the
hadronic phase.  The combusted phase is again $\beta$-stable.  The
vertical line corresponds to the central density of the most massive
stable configuration of a hyperonic star obtained using GM3 model
($\rho_h^{max}$=1.20 fm$^{-3}$).
\label{graf-beta-iperoni}}
}
\end{figure}

In Fig.~\ref{graf-nb} we show the results obtained for nucleonic
matter considering a not $\beta$-stable phase after the front, while
in Fig.~\ref{graf-beta} we show analogous results in the case of
$\beta$-stable matter.  Finally, in Fig.~\ref{graf-beta-iperoni} we
show the results for hyperonic matter assuming $\beta$-stability
immediately after the front.  In the lower panels of these figures the
pressure difference between the uncombusted and the combusted phase is
shown. Let us recall that $p_h-p_q>0$ corresponds to a subsonic
process and therefore to a deflagration (Eq.~\ref{pressure-condition}).
In the central panels we show the energy difference between the two
phases in the hadron phase rest frame. To have an exothermic process
this difference has to be positive, as discussed in
Sec.~\ref{temperature}.  It is also important to remark that
Eqs.~(\ref{ene},\ref{mom},\ref{baryon}) can admit multiple solutions,
but only one corresponds to an exothermic process.  Finally, in the
upper panels we show the difference between the velocity of the
combusted phase $v_q$ and the sound velocity $v_{sq}$ in the same
phase (all the velocities are in units of the velocity of light $c$).
As long as $v_q>v_{sq}$ a strong deflagration is obtained.  As it can
be seen, in all cases ($\beta$-stable and not $\beta$-stable one) the
conditions for detonation are not fulfilled\footnote{From a
quantitative viewpoint the conditions of
Eq.~(\ref{pressure-condition}) are more near to be satisfied if matter
after the conversion front is assumed to be not $\beta$-stable.
Comparing Figs.~\ref{graf-nb} and \ref{graf-beta} it is possible to
notice that in the case of $\beta$-stable matter $p_h>p_q$ for all
densities while in the not $\beta$-stable case there is a window of
low densities for which $p_q>p_h$ but there the process becomes
endothermic.  }.  In the upper panel of
Figs.~\ref{graf-beta},\ref{graf-beta-iperoni} and for the lowest value
of $B$ we indicate with a star the density below which quarks form
with a negative pressure, i.e. they are mechanically unstable
immediately after the conversion front.  From the central panel of the
same figures one can notice that the deconfinement transition stops
being exothermic at a density slightly larger and therefore quarks
produced in an real, exothermic process are indeed mechanically
stable.  On the other hand, from the central panel of
Figs.~\ref{graf-beta},\ref{graf-beta-iperoni} one can notice that for
$B^{1/4}=155$ MeV an exothermic process with a geometrical front is
possible only for very large densities, reachable only at the center
of very massive hadronic stars.  But the result of the conversion of
such massive hadronic star is not a compact star but a black hole, as
it can be seen from Fig.~\ref{fig-mr155}. Moreover massive hadronic
stars are presumably generated by mass accretion and therefore the
deconfinement process would likely take place when the mass of the
star is smaller.  Therefore we can conclude that for very small values
of $B$ the process of conversion is not associated with the formation
of a geometrical front.  Other conversion mechanisms can then take
place, for instance convection, and we will come back to this problem
in Sec.~\ref{convection}.

Up to now we have only considered zero temperature QM. On the other
hand, the process of thermalization of the newly formed phase takes
place through strong interaction and it is therefore possible that
immediately after the front matter has already to be considered as
thermalized. The new phase's temperature can be estimated from energy
conservation, as discussed in Sec.~\ref{temperature} and its value
reaches a maximum of a few ten MeV near the center of the star. In
Fig.~\ref{graf-nb-Tfinita} we show the results for a finite
temperature of the new phase.  Also in this case, a detonation is
never obtained.

\begin{figure}[ht]
\centering
\parbox{14cm}{
\scalebox{0.6}{
\includegraphics*[14,14][590,750]{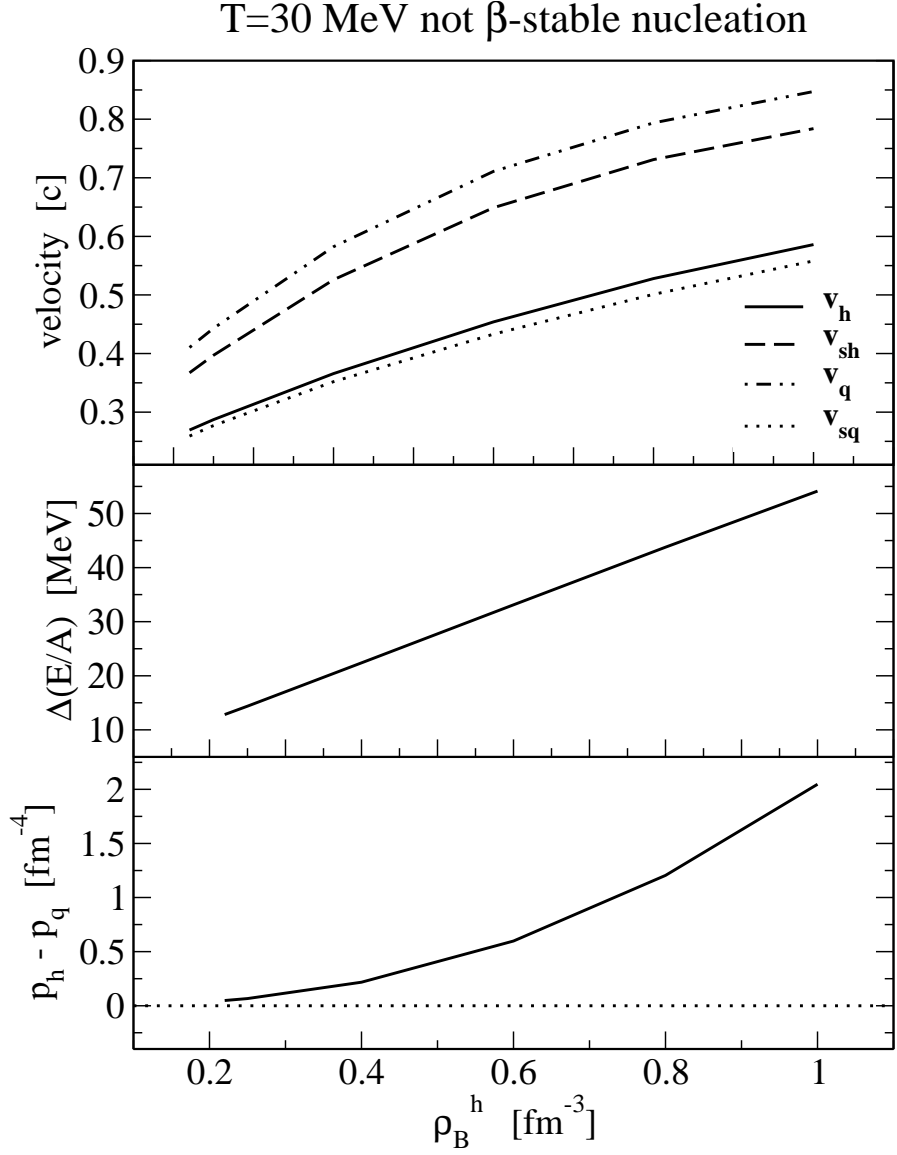}}
\caption{
\footnotesize
Notations as in Fig.~\ref{graf-nb-Tfinita}.
The combusted phase is obtained using $B^{1/4}=170$ MeV.
Here both the quark and the hadronic phase are at T$=30$ MeV.
\label{graf-nb-Tfinita-adroni}}
}
\end{figure}

We have also considered the possibility that heat transport
is more rapid than deconfinement and therefore hadronic matter near
the conversion front reaches a temperature similar to the one of newly
formed QM.  In Fig.~\ref{graf-nb-Tfinita-adroni} we show results for
this scenario and we conclude that also in this case the front is
unstable\footnote{Cho and Ng \citep{Cho93} found a little window of
densities in which detonation is possible but for rather unrealistic
high temperature.}.

\begin{figure}[!hb]
\centering
\parbox{14cm}{
\scalebox{0.6}{
\includegraphics*[14,14][590,750]{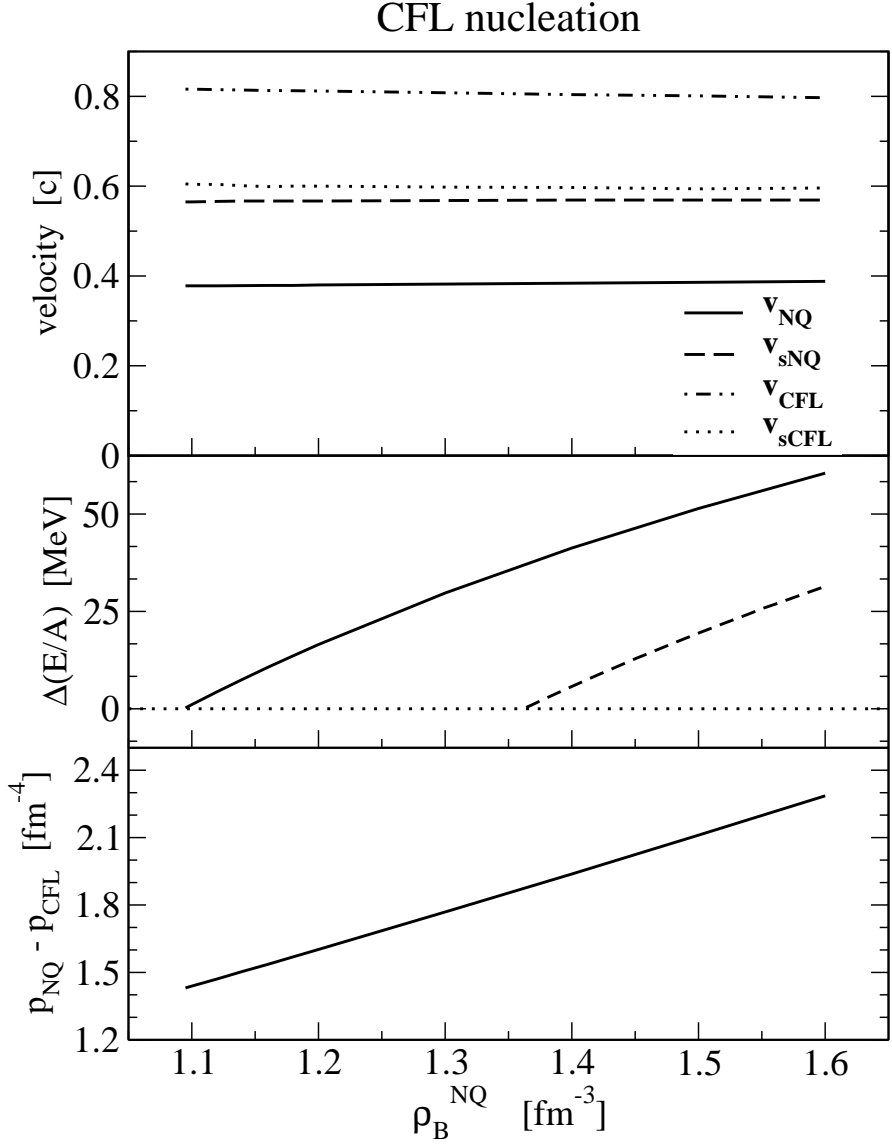}}
\caption{
\footnotesize 
Results for the conversion front separating $\beta$-stable quark
matter ($B^{1/4}=155$ MeV and $B^{1/4}=165$) from gapped quark matter
($\Delta=100$ MeV).  Upper panel: velocity of the Normal Quark phase
$v_\mathrm{NQ}$, of the diquark condensed CFL phase $v_\mathrm{CFL}$
and corresponding sound velocities $v_\mathrm{sNQ}$ and
$v_\mathrm{sCFL}$, all in units of the velocity of light and in the
front frame.  Here for simplicity only the results for $B^{1/4}=155$
are presented.  Center panel: energy difference between the two phases
(in the NQ phase rest frame).  Here we show the results both for
$B^{1/4}=155$ MeV (solid line) and for $B^{1/4}=165$ MeV (dashed
line).  Lower panel: pressure difference between the two phases. The
difference between the two pressures is almost exactly $B$
independent.
\label{graf-fronte2}}
}
\end{figure}

As discussed in Sec.~\ref{doublescenario}, it is possible that the
formation of diquark condensate takes place not immediately after the
deconfinement transition but it is delayed.  It is therefore
interesting to discuss also the transition from NQ to gapped QM, which
in our case we assume to be in a CFL phase.  In
Fig.~\ref{graf-fronte2} we show the results for this transition. Also
in this case no detonation is obtained and we are always in the regime
of strong deflagration.  The maximum central density for a stable
ungapped quark star is 1.411 fm$^{-3}$ for $B^{1/4}=155$ MeV and 1.619
fm$^{-3}$ for $B^{1/4}=165$ MeV. Therefore, at least for the central
region of stars near the maximum mass configuration a geometrical
front can indeed form.

Finally, it is important to remark that the velocity of sound in the
center of a compact star is typically of the order of $(0.5\div 0.8)
c$ and the velocity of the deflagrative front $v_{df}$ (defined in 
Sec.~\ref{fluid-eqs}) is marginally lower,
$(0.4\div 0.7) c$.

The main result obtained in this Section is that a detonative
regime is never directly reached after imposing the continuity conditions 
on the front. On the other hand two problems still need 
to be discussed: the estimate of the actual velocity of the 
deflagrative front taking into
account heat and strangeness diffusion and the effect of hydrodynamical
instabilities which in principle can increase the conversion velocity
transforming a deflagration into a detonation. These two points are 
discussed in the next Section.

\section{Hydrodynamic instabilities and effective velocity of the front 
\label{instability}}

In the previous section we have shown that the conversion process always 
takes place as a deflagration.
In this case it is extremely difficult to estimate the velocity of the 
conversion front. The velocity is governed by the slowest among the 
processes which need to take place for the combustion to continue.
In the seminal work of \citet{Olinto86} it was shown that, in the absence of 
hyperons, the conversion velocity crucially depends on the rapidity by 
which strangeness is produced in the quark sector and diffuses into the 
hadronic sector. 
The final expression for the velocity, assuming a stable laminar front 
(as if it was a slow combustion), depends on two quantities:
\begin{itemize}
\item the temperature of the system;
\item the down-strange asymmetry parameter
$a_0=(\bar\rho_d-\bar\rho_s)/2\, \rho_B$, which is related to the minimum 
number density of strange quarks $\bar\rho_s$ for which strange quark matter 
is absolutely stable.
Here $\bar\rho_d$ and $\bar\rho_s$ are the number densities of down and 
strange quarks, respectively.
\end{itemize}
Olinto only discussed the formation of a pure phase
of quark matter. Instead, in our paper we have also considered the 
possibility of producing a mixed phase of hadrons and quarks. It is 
actually possible to use the formalism of Olinto by re-interpreting 
the meaning of $a_0$ as the minimum strangeness content for which the 
conversion process is exothermic.
The velocity reads:
\be
v_{\mathrm{sc}}=\frac{2}{\sqrt{g}}\left( \frac{\mu}{T} \right) \, 
\mathrm{m}\, \mathrm{s^{-1}} \label{olinto}
\ee
where $g\simeq 2(1-a_0)/a_0^4$.
If strangeness indeed needs to be produced and diffused (i.e. $a_0\neq 1$)
the typical velocities estimated using Eq.~(\ref{olinto}) are rather low,
of the order of a few km/s for $T \sim$ 0.1 MeV.
The extreme case $a_0=1$ corresponds to 
the possibility of having an exothermic process in the absence of weak 
interactions.
In that situation the conversion is not delayed by strangeness diffusion and 
the actual velocity is limited by heat conduction, which was assumed as 
instantaneous in Olinto's analysis.
The characteristic timescale of heat diffusion (assuming a laminar front) 
reads:
\be
\label{heat-diffusion}
t_q \sim c_q\, l^2/k_q
\ee
where $c_q$ and $k_q$ are the specific heat and the heat conductivity of 
the quark phase and $l$ is the distance over which heat diffuses. The 
minimal distance for which Eq.~(\ref{heat-diffusion}) can be applied is 
of the order of the mean free path of the quarks $\lambda_q$\footnote{The 
quark mean free path is very large, up to 10$^3$ fm, due to
Pauli principle, see e.g. \citet{Olinto86}.}.
It is therefore possible to roughly estimate the thermalization timescale as
$\tau=t_q(l=\lambda_q)$ and, correspondingly, to define a heat diffusion 
velocity $v_{th}=\lambda_q/\tau$.
This estimate is similar to the one done by Olinto concerning the 
production of strangeness in the quark phase and its diffusion up to the 
hadronic phase.
Here heat is again produced in the quark phase and it has to propagate over 
a distance of the order of $\lambda$ to reach the hadronic phase.
Estimates of $c_q$ and $k_q$ can be found in the literature. In particular, 
for quarks exchanging perturbative gluons $c_q$ has been computed by 
\citet{Iwamoto82} and $k_q$ by \citet{Haensel:1989ja} and \citet{Baiko99}.
The heat diffusion velocity turns out to be of the order of a few percent 
of the velocity of light.
We are not computing the heat diffusion velocity in the case of CFL quark 
matter because there a laminar front cannot exist.

As already remark by \citet{Horvath88} the conversion velocity can be 
significantly increased taking into account hydrodynamical instabilities.
Indeed in the previous section we have shown that the conversion is always 
a strong deflagration and not a slow combustion. 
Therefore the conversion front is unstable and wrinkles can form.

There are at least two types of hydrodynamical instabilities 
discussed in literature, the Landau-Darrieus (LD) and the Rayleigh-Taylor 
(RT) (for an introduction to these problems see \citet{zeldovich85}).
Both these instabilities can develop when $\Delta e \equiv e_1-e_2>0$.
The LD instability is the one which characterizes the strong deflagration 
regime ($v_q>v_{sq}$, see Eq.~(\ref{strongdef})) and the amplification of the
wrinkles on the conversion front is directly due to the
conservation of the energy-momentum flux, as imposed
by Eqs.~(\ref{ene},\ref{mom},\ref{baryon}). 
That instability can develop independently
on the presence of gravity.
At the contrary, the RT instability develops
if a gravitational field is present and if the direction of the density
gradient is opposite to the direction of the gravitational force.

Due to RT and LD instabilities the area of the conversion front increases.
The conversion velocity also increases since all exchanges between the 
burned and the unburned zone are now more efficient.
A way of estimating the effective velocity $v_{\mathrm{eff}}$ is
through the introduction of the fractal dimension of the surface
\citep{woosley,blinnikov1,blinnikov2}. The larger is the
excess of the fractal dimension respect to the dimension
of a spherical front, the huger is the increase
of the front velocity respect to the laminar case.
In the absence of new dimensional scales between the minimal dimension 
$l_{\mathrm{min}}$ and the maximal dimension $l_{\mathrm{max}}$  of the 
wrinkle, $v_{\mathrm{eff}}$ is given by:
\be
v_{\mathrm{eff}}=v_{\mathrm{sc}} 
\left( \frac{l_{\mathrm{max}}}{l_{\mathrm{min}}} \right)^{D-2} \, .
\ee
Here $D$ is the fractal dimension of the surface of the front and it can 
be estimated as \citep{blinnikov1,blinnikov2}:
\be
D=2+D_0\,\gamma^2 \, ,
\ee 
where $D_0\sim 0.6$ and $\gamma = 1-e_2/e_1$.
In \citet{Lugones02} the effect of hydrodynamical instabilities on the 
conversion velocity has been discussed. Following their approach we start
by taking into account only the RT instability which are supposed to be 
the dominant ones.
In order to give a quantitative estimate of the velocity's increase, one needs 
to compute $l_{\mathrm{max}}$, $l_{\mathrm{min}}$ and $\gamma$.
Clearly enough $l_{\mathrm{max}}$ is of the order of a few Km.
Concerning $l_{\mathrm{min}}$ it can be estimated as the minimal size of the 
wrinkle for which the velocity of the RT growing modes is larger than 
$v_{\mathrm{sc}}$:
\be
l_{\mathrm{min}}=\frac{4\pi e_q \, v_{\mathrm{sc}}^2}{g \, \Delta e} \, ,
\ee
where g is the gravitational field:
\be
g(r) \equiv -\frac{1}{e(r)}\, \frac{dP}{dr} \, \label{grav}.
\ee
It is important to note that since $l_{\mathrm{min}}$ has to be smaller than 
$l_{\mathrm{max}}$ a maximum value for
$v_{\mathrm{sc}}$ can be defined. Laminar velocities exceeding 
$v_{\mathrm{sc}}^{\mathrm{max}}$ can not be further 
increased by RT instabilities.

\begin{figure}[!hb]
\centering
\parbox{14cm}{
\scalebox{0.6}{
\includegraphics*[14,14][590,750]{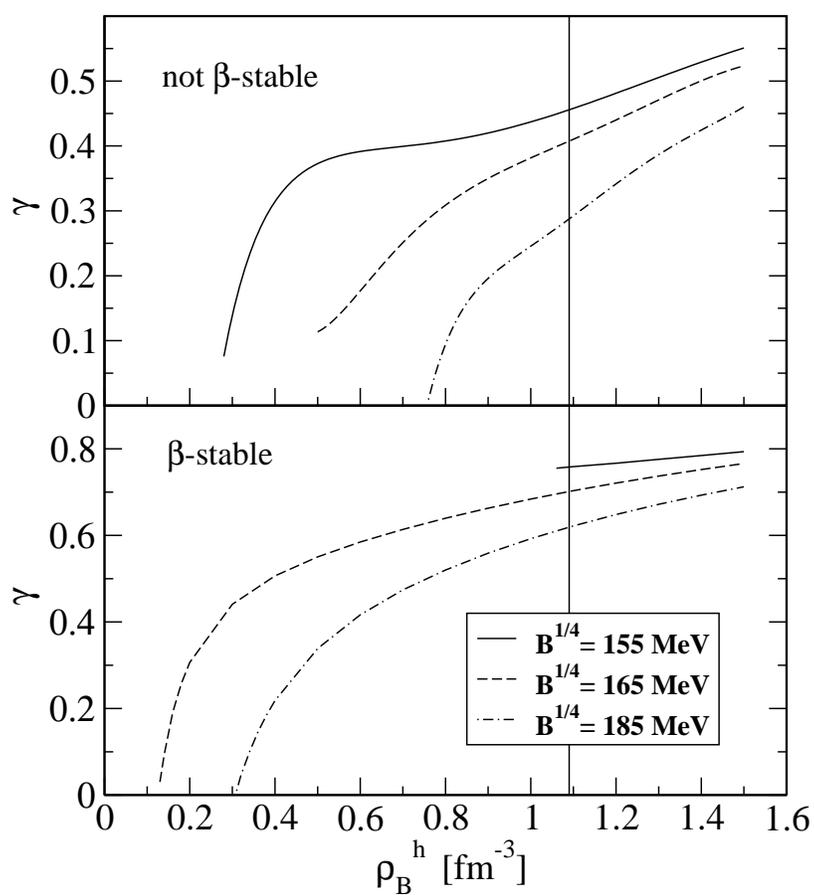}}
\caption{
\footnotesize The $\gamma$-factor entering the fractal dimension
of the conversion front, as a function of the baryonic density
of the hadronic phase.
The vertical line corresponds to the central density of the most
massive stable configuration of a nucleonic star obtained using GM3
model ($\rho_h^{max}$=1.09 fm$^{-3}$).
\label{gamma}}
}
\end{figure}

In Fig.~\ref{gamma} we show the values of $\gamma$ as a function of the 
density for various values of $B$ both in the $\beta$-stable and in the not 
$\beta$-stable scenario. In the not $\beta$-stable case the maximum value of 
$D-2$ is $\sim 0.12$ (for $\gamma \sim 0.45$). 
Instead, in the $\beta$-stable case, $D-2$ can be as large as $\sim 0.34$.

As an example, let us consider an hadronic star with $M_G=1.4 M_\odot$ 
for which the maximum density is 0.51 fm$^{-3}$.
For $B^{1/4}=155$ MeV and in the not $\beta$-stable case we obtain at 
the center of the star
$v_{\mathrm{eff}}= 0.48 \, v_{\mathrm{sc}}^{0.83}$. In this case 
$v_{\mathrm{sc}}^{\mathrm{max}}=0.056$ c.
For $B^{1/4}=165$ MeV and in the $\beta$-stable case we obtain at 
the center of the star
$v_{\mathrm{eff}}= 0.40 \, v_{\mathrm{sc}}^{0.64}$. In this case 
$v_{\mathrm{sc}}^{\mathrm{max}}=0.079$ c.
It is clear that RT instability alone cannot increase the conversion velocity 
above the velocity of sound and therefore they cannot transform a 
deflagration in a detonation.

It is well known that LD instabilities significantly increase the
velocity of the conversion. To estimate their effect one can again
resort to the fractal scheme, by substituting $l_{\mathrm{min}}$ with a 
new minimal scale $l_\mathrm{crit}$ below which LD instabilities are 
suppressed.
For instance, in the case of Supernovae type Ia several studies indicate
that $l_\mathrm{crit}\sim 100 \,l_\mathrm{th}$, where $l_\mathrm{th}$
is the thickness of the flame (see e.g. \citet{woosley97}).
If that scheme can be applied also to quark deconfinement,
$l_\mathrm{crit}\sim 10^5$ fm, since it seems reasonable to assume
$l_\mathrm{th}\sim\lambda_q\sim 10^3$ fm. Although these estimates are 
clearly very uncertain, taking into account the smallness of $\gamma$ 
in the not $\beta$-stable case, the velocity can increase by less than 
one order of magnitude. The conversion process should therefore remain 
subsonic but for very special choices of the equation of state parameters. 
In the $\beta$-stable case the velocity can increase by maybe two orders 
of magnitude, since the $\gamma$ is larger. 
On the other hand in that case $v_{\mathrm{sc}}$ is extremely small,
as estimated from Eq.~(\ref{olinto}). Therefore in the $\beta$-stable case 
it is extremely unlikely that the the hydrodynamical instabilities can 
transform the deflagration into a detonation. Nevertheless they increase 
the conversion velocity by several orders of magnitude, what is important 
in astrophysical applications.

\section{Convection}

\label{convection}

\subsection{Conditions for the existence of a convective layer}

In the previous Section we have discussed instabilities which can
increase the conversion velocity by forming wrinkles on the 
front surface. Another way of accelerating the burning process
is via the formation of a convective layer above the conversion front.
Here we discuss under which conditions convection can develop
during the process of quark deconfinement.

First let us recall that the energy density of the newly formed
QM just after the conversion front is {\it smaller} 
than the energy density of hadronic matter immediately before the front.
Therefore blobs of QM can try to penetrate the unburned
hadronic matter and in principle convection could instaure.
For convection to actually develop, the previous condition
is necessary but not sufficient, because the drop of QM is formed at a 
pressure which is also {\it smaller} than the pressure of hadronic
matter \citep{landau,Horvath88}.
The condition for convection to develop reads:
\be
e(P_B,S_B,Y_e^B) < e(P,S,Y_e)\, , \label{convect}
\ee 
where $S$ is the entropy and $Y_e$
is electron fraction and the suffix $B$ indicates the same quantities
for the blob.
It is called {\it mixing length} the distance traveled by the blob
before being so modified by the surrounding medium that condition
(\ref{convect}) is no more satisfied.  

There are various types of
convection.  In order for the so-called ``quasi-Ledoux'' convection to
develop, the inequality of Eq.~(\ref{convect}) has to be satisfied
with $P_B=P$, in every point of the convection layer, whose size is
actually defined through Eq.~(\ref{convect}) itself (for a recent
review of hydrodynamical problems see e.g. \citet{Wilson2003}).
More explicitly, 
as soon as the quarks' drop enters the hadronic phase, the pressure of
the blob starts equilibrating with the pressure of the surrounding
material.  If the inequality of Eq.~(\ref{convect}) with $P_B=P$ is not
satisfied, the quasi-Ledoux convection can not develop.  

A slightly
more general type of convection is associated with the 
question: how long can the quarks' drop travel inside hadronic matter
before the pressure equilibrates?  The answer to this question depends
mainly on the size of the drop $R_{B}$ because the number of 
scatterings, due to strong
interaction and needed to equilibrate the pressure is of the
order of the baryon number of the drop.  Therefore the mixing length
is also of the order of $R_{B}$.  As we will show in the next
subsection, using realistic EoSs, the 
most relevant case in which
quasi-Ledoux condition is not
satisfied corresponds to the situation in which mixed phase is produced. 
A natural length scale for $R_B$ can then be the size of the structures 
of the mixed phase, which is typically of the order of a few fermis
\citep{Heiselberg:1992dx} and this is also the distance traveled by
the drop before strong interactions push the drop back into the quark
phase. In conclusion, the quasi-Ledoux convection is the only one we
will discuss in the following, because other possible convection
mechanisms are suppressed in the system we are discussing here.

In our calculation hadronic matter is taken to be cold while the newly
formed quarks are in principle at a finite temperature, which can be
estimated as done in Sec.~\ref{temperature}.  Anyway, in the case of
massless quarks a finite temperature plays no role, because the
relation between density of energy and pressure is independent on the
temperature in the massless limit.  Taking into account the finite
value of the strange quark mass, the energy-pressure relation is
temperature dependent, but we have checked that the effect of the
temperature on convection is totally negligible, as it will be
clarified in the next subsection\footnote{In principle one has also to
use an isoentropic EoS to describe the evolution of the structure of
the drop inside the hadronic phase. Also in this case we have checked
that the effect of the temperature on the energy density of the drop
is totally negligible.}.  The system we are discussing is rather
different from other system in which convection develops due to the
finite temperature of the drop. In our case the system is strongly
degenerate and temperature plays only a minor role. Therefore the size
of the mixing-length in our case is not determined by the heath
dissipation of the drop, but only by the time needed for strong
interactions to equilibrate the pressure and energy density, squeezing
the drop.

\subsection{Convection for realistic EoSs}
\label{realistic-convection}

\begin{figure}[!ht]
\begin{center}
\includegraphics*[scale=0.46]{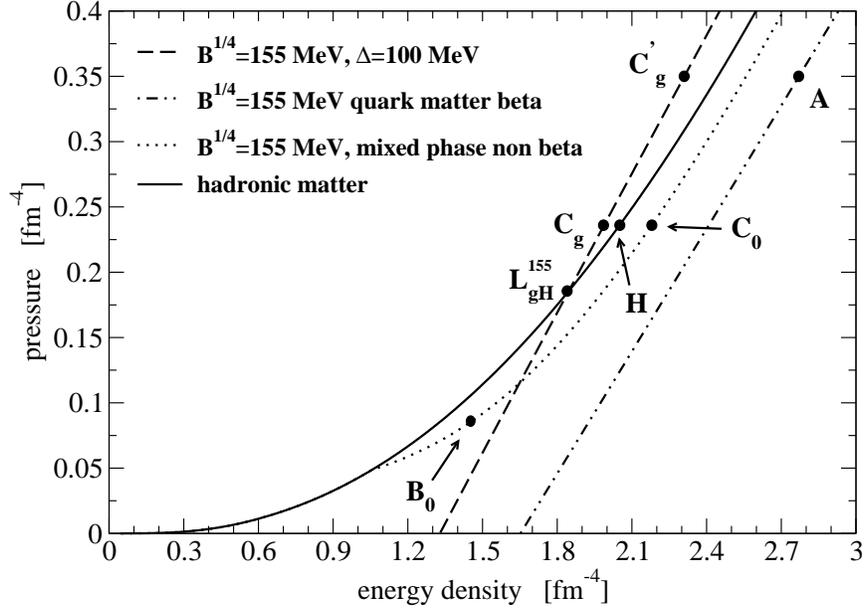}
\end{center}
\parbox{14cm}{
\caption{
\footnotesize 
Scheme for convection: H represents the drop of hadronic matter just
before deconfinement, B$_0$ represents the drop of newly formed QM, C
stays for the drop of QM after pressure equilibration and L indicates
the end point of the convective layer. Finally A represents a drop of
ungapped quark matter before its transition to CFL phase.  Here
$B^{1/4}=155$ MeV and hyperons are not included.
\label{fig-convection155}}
}
\end{figure}

\begin{figure}[!hb]
\begin{center}
\includegraphics*[scale=0.46]{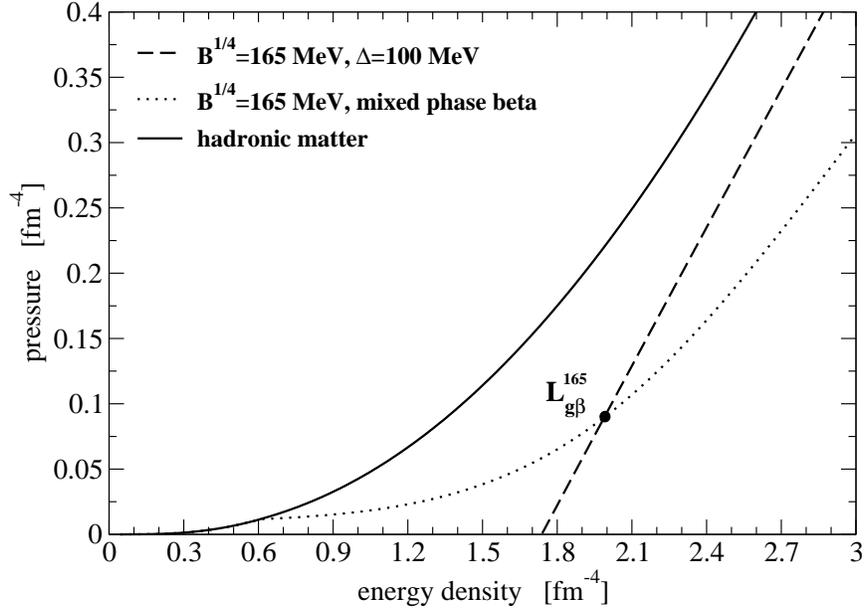}
\end{center}
\parbox{14cm}{
\caption{
\footnotesize Scheme for convection: see Fig.~\ref{fig-convection155}.
Here $B^{1/4}=165$ MeV and hyperons are not included.
\label{fig-convection165}}
}
\end{figure}

\begin{figure}[!ht]
\begin{center}
\includegraphics*[scale=0.47]{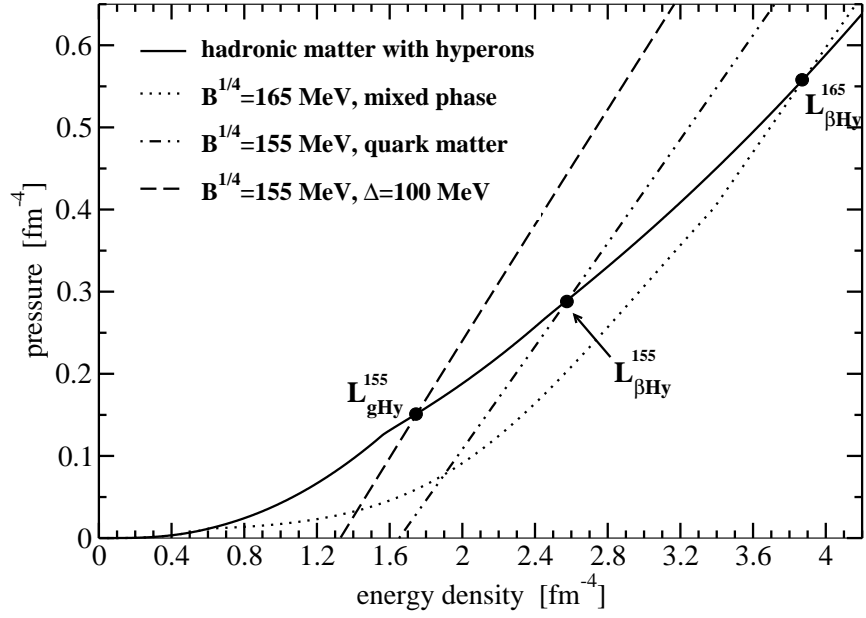}
\end{center}
\parbox{14cm}{
\caption{
\footnotesize 
Scheme for convection: see Fig.~\ref{fig-convection155}.
Here $B^{1/4}=155$ MeV and 165 MeV and hyperons are included.
\label{fig-convection-iperoni}}
}
\end{figure}

In
Figs.~\ref{fig-convection155},\ref{fig-convection165},\ref{fig-convection-iperoni}
we show the results of the analysis of the quasi-Ledoux convection
using the EoSs and the compact star profiles discussed in previous
sections.  In Fig.~\ref{fig-convection155} (where $B^{1/4}$ = 155 MeV)
we also show examples of trajectories in the pressure vs
energy-density plane of a drop of quarks after its formation.  We have
two sets of letters describing the trajectory of the quark drop,
namely the one corresponding to ungapped and not $\beta$-stable mixed
phase (suffix 0) and letters corresponding to CFL quark matter, with
suffix $g$.  In principle we should also have letters with suffix
$\beta$ corresponding to ungapped $\beta$-stable QM, but as discussed
in Sec.~\ref{deflagration}, in that case it is not possible to form a
geometrical conversion front for $B^{1/4}$ = 155 MeV.  The original
drop of hadronic matter from which quarks have formed is indicated
with H. With B$_0$ we indicate the energy and pressure of the quarks
immediately after deconfinement. As discussed in
Sec.~\ref{deflagration} 
the pressure and energy density of quarks are
smaller than those of hadrons, in agreement with the characteristics
of an unstable deflagration front. Since the density of the quark drop
is smaller than the density of the hadronic medium the drop will start
moving in the opposite direction of gravity, but, as soon as the drop
enters hadronic matter its pressure has to equilibrate with that of
hadrons (C$_0$).  It is clear that in the case of not $\beta$-stable
mixed phase convection cannot develop, because while $e ($B$_0) < e
($H$)$, after pressure equilibration $e($C$_0) > e($H$)$ and therefore
the drop, immediately after entering hadronic matter, has an energy
density {\it larger} than the hadronic energy density in the same
position and the drop is forced to sink back into the QM region. As
noted above, the temperature of the quark phase plays no role here
because the quark line in the energy-pressure plane is almost
independent on the temperature.

From Figs.~\ref{fig-convection155},\ref{fig-convection165} we notice
that convection can never develop if hyperons are not present,
independent on the value of $B$, as long as diquark condensate is not
formed.  Fig.~\ref{fig-convection-iperoni} shows that convection can
develop if hyperons are present both for $B^{1/4}=155$ MeV and
$B^{1/4}=165$ MeV.  Convection can not develop for $B^{1/4}=185$ MeV.

In Fig.~\ref{fig-convection155} we also show the effect of the
formation of a diquark condensate. The scheme we have in mind is the
following:
\begin{itemize}
\item the transition from hadronic matter to ungapped QM takes place
as described above. The velocity of the conversion front is not very
large because convection cannot develop;
\item in a random site inside the already formed NQ quarks start
gapping.  Notice that the formation of a CFL gapped quark drop is
delayed due to the need to reach $\beta$-stability (producing
strangeness and equilibrating the up and down quark content), to
deleptonize and to allow the cooling of the star.  Moreover, the
transition from ungapped to CFL quark matter appear to be first order
\citep{Ruster:2005ib} and therefore a finite nucleation time has to be
taken into account;
\item the conversion from ungapped to CFL quarks proceeds as a 
deflagration, as results from the analysis presented in
Fig.~\ref{graf-fronte2}, but in this case {\it convection can
develop}.  This is clear if one considers a drop of $\beta$-stable NQ
(A) which transforms into a drop of CFL quarks and then equilibrate
its pressure reaching the point C'$_g$, with $e ($C'$_g) < e ($A$)$;
\item due to convection, the conversion front separating NQ from CFL
phase moves rapidly outwards;
\item if the time delay in formation of the first drop of CFL quarks
is not too large, the new conversion front can reach the ``slow''
front separating hadronic matter from NQ.  As it is clear from
Fig.~\ref{fig-convection155}, drops of CFL quarks can penetrate
hadronic matter, generating convection even inside hadronic matter (as
long as the energy of the blob $e_b > e(\mathrm{L_{gH}^{155}})$~) and
making the full conversion process considerably faster. Here and in
the following we indicate with L the endpoint of the convective layer.
The lower indexes indicate the phase of which the drop is made and the
phase in which it propagates, while the upper index indicates the
value of $B^{1/4}$.\\ While the existence of a convective layer during
the conversion of unpaired QM into gapped QM is independent of the
model parameters, the possibility that this convective layer extends
inside the hadronic matter region strongly depends on the specific
value of the parameters.
\end{itemize} 

We discuss now the thickness of the superconducting layer $\lambda_c$
and the velocity $v_c$ by which convection expands.  Concerning the
first point, it is clear that in our scheme the convective layer
extends from the center of the star down to the layer whose energy
density is $e($L$)$. Typically, $\lambda_c$ is of order of a few km.
In Figs.~\ref{fig-pro155},~\ref{fig-pro165} we show the endpoints of
the convective layers\footnote{For $B^{1/4}=185$ MeV convection can
not develop, as shown in Fig.~\ref{fig-pro185}.}.  With NQ we indicate
the end of the layer where drops of ungapped $\beta$-stable quarks can
develop convection.  Similarly we indicate with CFL the layer in which
gapped quarks can develop convection.

Concerning the velocity $v_c$ of expansion of the convective front,
it can be estimated from the conservation of total energy as:
\be
\label{eq-convec}
\frac{1}{2} e_b(\mathrm{L})\, \mathcal{V}_f\, v_c^2 = 
e_b(\mathrm{C})\,  \mathcal{V}_i\, U[r(\mathrm{C})]-
e_b(\mathrm{L})\,  \mathcal{V}_f\, U[r(\mathrm{L})]
\ee
where $U(r)$ if the gravity potential and e$_b$(C) is the energy
density of the blob when its pressure equals the pressure of the
surrounding medium and the blob starts being accelerated by the
bouyant forces.  The initial and final values of the volume $\mathcal{V}$ 
of the drop of QM are related by baryon number conservation:
\be
\mathcal{V}_f \,\rho_B(\mathrm{L}) = \mathcal{V}_i \,\rho_B(\mathrm{C}) \, .
\ee
A simpler equation for $v_c$ can be obtained by expanding
Eq.~(\ref{eq-convec}) up to first order in the energy density
difference $\delta e_b$.  Following \citet{Wilson2003} the equation
for $v_c$ can then be written as:
\be
\label{eq-convec-simple}
\frac{1}{2} \, e_b \, v_c^2=\delta e_b \, g \, R_c 
\ee
where $R_c$ is the distance traveled by the blob, 
i.e. the distance from where it has been created up to 
the end of the convective layer. The gravitational field $g$ is defined in 
Eq.~(\ref{grav}).

Using Eq.~(\ref{eq-convec-simple}) we obtained values of the convective
velocity ranging from $\sim 10^4$ km/s when the drop is produced at
relatively low densities to $\sim 10^5$ km/s when it is produced near
the center of a massive star.
Although the convective velocity
is large, it is anyway much lower than the sound velocity.

\section{Conclusions and astrophysical implications}
\label{results}

To clarify the astrophysical implications of our work let us discuss 
two scenarios for the formation of QM in which our
formalism can be applied.  

In the first scenario the surface tension
at the interface between hadrons and quarks is negligible or vanishes, 
i.e. the star can not become metastable respect to the formation of 
a drop of QM ($\sigma$ smaller then a few MeV/fm$^2$).  In this case 
the formation of QM takes place soon after the supernova explosion. In
agreement with the analysis of \citet{Pons:2001ar}, QM starts forming
when the proto-neutron star has deleptonized and its temperature drops
down to a few MeV.  At the center of the proto-neutron star a drop of
QM can form and, if $B$ is not too small, a finite strangeness content
is needed for the deconfinement process to be exothermic. The
strangeness content could be already present in the hadronic star due
to the formation of hyperons.  When the drop starts expanding, the
process of conversion can be extremely fast within the layer
in which deconfinement is energetically convenient even in the absence
of weak processes.  In this case the conversion front moves at the
velocity of the deflagrative front $v_{\mathrm{df}}$, which approaches 
the velocity of sound (see Sec.~\ref{deflagration}). 
As shown in
Figs.~\ref{fig-eos155}, \ref{fig-eos165}, only at very large densities
the EoS of not $\beta$-stable matter is composed of pure quarks.
As the conversion layer moves outward, the front enters the region of 
mixed phase where $v_{\mathrm{df}}$ decreases till it vanishes at the low 
density boundary of the mixed phase. Before reaching that point the 
conversion process involving weak reactions becomes first competitive 
and then dominant.
The diffusion of strangeness is a relatively slow process whose velocity 
can nevertheless be significantly increased by hydrodynamical instabilities.
In this scenario heat diffusion plays a marginal role.

In the alternative scenario the surface tension is
larger and the hadronic star can therefore become metastable. 
In the region in which pure quark matter can form there is no
substantial difference with the previous scenario
although heat diffusion can be necessary to have a rapid expansion
of the pure QM (see discussion at the end of Sec.~\ref{tn}). 
When the region of mixed
phase is reached, the only way of rapidly producing this new
phase is via thermal nucleation. In this scenario the deflagrative
velocity is therefore limited by the heat diffusion velocity
if strangeness need not to diffuse. Again strangeness diffusion
will be crucial to convert the outer layers of the star.

If the surface tension exceeds few ten MeV/fm$^2$ the process of formation
of mixed phase is extremely slow and it cannot be described using a  
fluidodynamical scheme.

The main result of our analysis, based on realistic EoSs, is that the
conversion from hadronic matter to QM, or to a mixed phase of hadrons
and quarks and also the transition from unpaired QM to gapped QM
always takes place as a deflagration.  This result does not
change if a finite temperature of the system is taken into account. In
our analysis we have shown that the maximum temperature obtained is
$\sim$ 50 MeV near the center of a massive star.  For such a
relatively low temperature the system remains strongly degenerate and
thermal effects are small.  
To estimate the increase of the conversion velocity
due to hydrodynamical instabilities we have used a 
fractal scheme. Although the wrinkles which develop on the front surface
can significantly increase the conversion velocity,
in most realistic cases the process remains subsonic and the
transformation from deflagration to detonation does not take place.
Concerning the possibility of developing
convection, this is possible if hyperons are present and if $B$ is not
too large and the mass of the compact star not too small. Convection
can also develop if quarks can form a condensate. In particular, in the
conversion from ungapped to gapped QM convection always takes
place.

Let us now discuss two astrophysical problems in which the 
type of conversion, either deflagrative or detonative, and the 
conversion velocity play a crucial role.

\subsubsection*{Neutron star velocities}
It has been proposed by \citet{Bombaci:2004nu} that the high
velocities displayed by some neutron stars can be attributed to an
asymmetric neutrino emission associated with the formation of QM
inside the hadronic star.  The origin of this asymmetry could be
related to a process of deconfinement starting off the center of the
star.  

In our analysis we have shown that if hyperons are present or
if a diquark condensate forms then convection can develop.
The possibility of rapidly transporting hot material to the 
surface of the star via the formation of a convective layer
can indeed 
be at the origin of strong asymmetries in the conversion 
process\footnote{The model proposed in \citet{Bombaci:2004nu} and here 
discussed has {\it no connection} with models in which the kicks are 
explained as due to parity violating processes in the presence of a 
strong magnetic field, a mechanism which is known to provide almost no 
contribution to the neutron star velocity.}.

\subsubsection*{Gamma Ray Bursts}
It has been speculated several times
\citep{Cheng:1995am,Bombaci:2000cv,Wang:1999mf,Ouyed:2001cg,
Berezhiani:2002ks,Bombaci:2004mt, vidana:2005} that the so-called long
GRBs can be originated by the conversion of an hadronic star into a
star containing deconfined QM, either as a pure phase or in phase in
which quarks are mixed with hadrons. Moreover, as discussed in
\citet{Drago:2005qb} (see Sec.~\ref{doublescenario} and the analysis
of the time-structure of the light curves of GRBs presented in
\citet{Drago:2005cc}) the conversion process can take place in two
steps, with a first transition from hadrons to ungapped (or 2SC)
quarks and a second transition in which a CFL phase is produced. In
order to associate an emission peak with each of the two transitions,
the conversion process must be rapid enough to deposit in a few
seconds (or less) a huge energy inside the star. Neutrinos will then
transport the energy to the exterior on a time scale of order $(10\div
20)$~s. Clearly, the result of our calculation provides these large
velocities, since the conversion process occurs on a time scale of
$(0.1\div 1)$ s for the first transition in the case of a 
laminar front and it is much more rapid if the hydrodynamical
instabilities are taken into account. The second transition
lasts only some 10$^{-3}$ s due to the formation of
a convective layer. If the two processes takes place one after
the other it is even possible that the formation of diquark condensate
accelerates the conversion process by developing a convective layer
inside the hadronic phase.

\noindent
It is also important to recall that
the way in which the conversion to quark matter takes place,
either via a detonation or a deflagration, is crucial.
It has been shown that the mechanical wave associated with a 
detonation would expel a relatively large amount of baryon from the
star surface \citep{fryer}. In the case of a detonation the region
near the surface of the compact star where the electron-photon
plasma forms (via neutrino-antineutrino annihilation) would be 
contaminated by the baryonic load and it would be impossible to 
accelerate the plasma up to the enormous Lorentz factors needed to 
explain the GRBs.

\section{Acknowledgments}

It is a pleasure to thank S. Blinnikov, G. Pagliara and R. Tripiccione 
for many useful discussions.

\newpage

%%%% figures %%%%%%%%%%%%%%%%%%

\newpage

\newpage

\newpage

\newpage

\newpage

\newpage

\newpage
\clearpage

\newpage

\newpage

\newpage

\newpage

\newpage

\newpage

\newpage

\newpage
\clearpage

\end{document}